\documentclass[runningheads]{llncs} 
\usepackage{lscape}
\usepackage{url}
\usepackage{longtable}
\usepackage{float}
\usepackage{multirow}
\usepackage{amsmath}
\usepackage{graphicx}
\usepackage{subfigure}
\usepackage{amssymb}
\usepackage{lineno}
\usepackage{booktabs}
\usepackage{textcomp}
\usepackage{threeparttable}

\begin{document}

\begin{frontmatter}

\title{Effects of Regional Trade Agreement to Local and Global Trade Purity Relationships\thanks{Supported by the Chinese National Natural Science Foundation (71701018, 61673070) and the National Social Sciences Fund, China (14BSH024).
}}

\author{Siyu Huang\inst{1},Wensha Gou\inst{1}, Hongbo Cai\inst{3}\\ Xiaomeng Li\inst{1}* and  
Qinghua Chen\inst{1,2}*}
\date{}

\institute{School of Systems Science, Beijing Normal University, Beijing, China \\ \and New England Complex Systems Institute, Cambridge, MA, USA\\  \and Business School, Beijing Normal University, Beijing, China\\
*corresponding author: \email{lixiaomeng@bnu.edu.cn,qinghuachen@bnu.edu.cn}}

\date{June. 2019}


\authorrunning{S. Huang et al.}
\maketitle    

\begin{abstract}

In contrast to the rapid integration of the world economy, many regional trade agreements (RTAs) have also emerged since the early 1990s. This seeming contradiction has encouraged scholars and policy makers to explore the true effects of RTAs, including both regional and global trade relationships. This paper defines synthesized trade resistance and decomposes it into natural and artificial factors. Here, we separate the influence of geographical distance, economic volume, overall increases in transportation and labor costs and use the expectation maximization algorithm to optimize the parameters and quantify the trade purity indicator, which describes the true global trade environment and relationships among countries. This indicates that although global and most regional trade relations gradually deteriorated during the period 2007-2017, RTAs generate trade relations among members, especially contributing to the relative prosperity of EU and NAFTA countries. In addition, we apply the network to reflect the purity of the trade relations among countries. The effects of RTAs can be analyzed by comparing typical trade unions and trade communities, which are presented using an empirical network structure. This analysis shows that the community structure is quite consistent with some trade unions, and 
the representative RTAs constitute the core structure of international trade network. However, the role of trade unions has weakened, and multilateral trade liberalization has accelerated in the past decade. This means that more countries have recently tended to expand their trading partners outside of these unions rather than limit their trading activities to RTAs.

\end{abstract}
\end{frontmatter}

\clearpage

\section{Introduction}

With the rapid development of international trade, as of 2020, the World Trade Organization (WTO) has 164 members representing 98 percent of world trade. However, in addition to this extensive multilateral trading system, the world has also witnessed unprecedented proliferation of regional trade agreements (RTAs) since the 1990s~\cite{Bao2019The}. In 2013, 546 notifications of RTAs were received by the General Agreement on Tariffs and Trade (GATT)/WTO~\cite{Dai2014}. The role of RTAs raises questions among scholars and policy makers: what drives an increasing number of countries to join regional trade unions, and how will this affect regional trade patterns and globalization processes? Trade creation and trade diversion have been proposed to describe the effects of RTAs~\cite{Carr2006Revisiting,RePEc:ags:eprcrs:166082}. Trade creation refers to new trade arising between member countries due to the deduction of tariffs, while trade diversion means that imports from a low-cost outsider country are replaced by imports from a higher cost member country because of RTA~\cite{Yang2014}. Some have advocated for RTAs by arguing that, unlike multilateral trade liberalization, they promote``deeper" integration~\cite{bhagwati1996the}.

Despite the controversy in the literature, previous studies usually focus on the influence of RTAs on countries in given regions instead of quantitative analyses on a global scale. A common approach is to operationalize RTA membership as a categorical independent variable and analyze the influence of trade unions on bilateral trade using a gravity model~\cite{Vicard2009On,Reyes2014,Ghosh2004Are,jung2008gravity}. However, the roles of RTAs in regional and global trade differ, which can also be seen in the description of trade creation and trade diversion. It is not comprehensive to study them separately, and we need to break through the limitations of existing research. In fact, international trade is a complex system with global characteristics and regional structures, and we should analyze the effects of RTAs on both regional and global trade environments. It is necessary to use quantitative models and network methods to analyze global trade as a whole, and the influence of other countries should not be ignored when discussing the trade flow between any two countries.

RTAs are usually signed between neighboring countries, so their effects on regional trade are coupled with geographical distance and other factors. The innovation of this paper is to study and describe the trade purity relationship of countries, with some other typical factors, such as economic volume, geographical distance, overall increases in transportation and labor costs, are separated. In contrast to the existing literature, which consistently increases observable variables to quantify trade costs~\cite{Deardorff2004Local,Tadesse2017Does,Liu2018}, here, we define synthesized trade resistance~\cite{Anderson2010The}, decompose it into natural and artificial factors, and propose a trade purity indicator (TPI) to describe the true trade environment and relationships between countries. The role of RTAs can be studied by comparing the TPI and its evolution within and outside a trade union. Here, we apply the expectation maximization (EM) algorithm to optimize the parameters and quantify the trade purity indicator.
Compared with the exogenous parameter estimation in the existing research on trade cost quantification~\cite{anderson2004,helpman2008estimating,Chaney2008Distorted,corcos2011}, the method in this paper is more scientific and effective, and it could be extended to discuss the effects of RTAs on a number of countries around the world.

Furthermore, international trade is a system that involves numerous countries and trade relations, and complex network modeling has the advantage of analyzing a number of entities and complex relationships~\cite{Song2017Topological,Zhang2019,Zhang2006Network}. Additionally, network theory can also facilitate the examination of both local and global properties~\cite{Zhong2014The}, which is consistent with the goal of our work. However, trade flows are a direct result of trade openness, and related studies usually apply trade flows to weight the network~\cite{Song2017Topological,Reyes2014}. Since trade flows could be influenced by a country's economic volume, geographical factors and artificial barriers, we prefer trade resistance, which removes the impact of the economy, to reflect the purity of the trade relationship between countries. In addition, communities in the international trade network are represented by clusters of countries where trade relations between countries in the same community are closer than those in different communities~\cite{Reyes2014}. Therefore, comparing the members of typical trade unions and trade communities in the global trade network could facilitate research on the effectiveness of RTAs.

The paper is organized as follows: Section 2 briefly describes the data source and the gravity model with synthesized trade resistance. Here, we establish a maximum likelihood function to simultaneously estimate the unobserved parameters and quantify the trade purity indicator. Section 3 presents the results. Here, we focus on six typical RTAs: Belt and Road (BRI), European Union (EU), North American Free Trade Agreement (NAFTA), Organization of African Union (OAU), Caribbean Free Trade Area (CARIFTA), and Association of Southeast Asian Nations (ASEAN). We discuss the evolution of TPI at both the regional and global trade levels and analyze the effects of RTAs during the period 2007-2017. In addition, we discuss the evolution of trade communities based on network methods. This shows that the representative RTAs constitute the core structure of international trade network, but the role of trade unions has weakened and multilateral trade liberalization has accelerated in the past decade. Finally, Section 4 provides the conclusion and discussion.

\section{Data and Methods}

\subsection{Data Source}

In this paper, we use trade data from the UN Comtrade Database, which includes 198 countries/districts. Here, we choose the ``Goods'' type of product and use the annual total of all Hs commodities (Harmonized Commodity Description and Coding Systems). In view of differences in time and statistical caliber, the flow data reported by the importer and exporter are not always the same. Here, we use the importer's report, with a supplement from exporters when the data are missing.

\begin{table}[tb]
	\centering 
	\caption{Data Description}  
	\label{table1} 
	\scriptsize
\begin{tabular}{|c|p{5.3cm}|p{5.5cm}|}  
Indicator & Indicator Description & Data Source\\
\hline
Trade Flows & Country-to-Country Trade Flows, from UN Comtrade Database. For `Goods', `Hs' commodities, annual data during the period 2007-2017. & \url{https://comtrade.un.org/}\\ \hline
GDP & GDP (current US\$) for countries, from the World Bank database, with code NY.GDP.MKTP.CD., annual data during the period 2007-2017. & \url{https://data.worldbank.org/indicator/NY.GDP.MKTP.CD}\\ 
\hline
Distance &	Geographical distance between mean positions of countries. The coordinate data are from Blue Marble Geographics. & \url{https://www.bluemarblegeo.com/index.php}\\ \hline
\end{tabular}
\end{table}

For GDP (current US\$), we use the World Bank national accounts data and OECD National Accounts data files. It is calculated without making deductions for depreciation of fabricated assets or for the depletion and degradation of natural resources. Data are in current US dollars. Dollar figures for GDP are converted from domestic currencies using single-year official exchange rates.

There are several methods for calculating geographical distance. As some countries have many import and export ports, we do not choose the coordinates of the capital but use the mean position of the longitude and latitude to calculate the distance. A full description of the data sources is provided in Table \ref{table1}.

\subsection{Quantifying Trade Resistance with a Gravity Model}

The gravity model is one of the most successful empirical methods in the field of social science~\cite{Anderson2010The}. Specifically, Isard and Tinbergen were pioneers in applying the gravity model to describe the patterns of bilateral aggregate trade flows among countries~\cite{isard1954location,tinbergen1963shaping}. Their work spawned a vast empirical literature that appears to perform well at modeling trade flows and exploring the factors influencing them~\cite{Anderson2010The,Head2014Chapter,Krisztin2015}, as $80\% - 90\%$ of the variation in the flows could be captured by the fitted relationship~\cite{anderson2011gravity}.

Scholars have introduce possible explanatory variables and performed regressions with panel data to confirm whether trade growth or loss is more significant~\cite{Yang2014,Cipollina2010Reciprocal,Carr2006Revisiting,Anderson2003}. However, it is impossible to include all the relevant factors, so the estimation of effects might be biased and inconsistent due to omitted variables, with the possibility of significant over- or underestimation~\cite{Reyes2014}.

In Tinbergen's gravity model, distance $d_{i,j}$ is not limited to geographical distance, and it could be broadly construed to include all factors that might create trade resistance~\cite{tinbergen1963shaping,Krisztin2015}. More recently, some papers have estimated synthesized trade costs or resistance from the observed pattern of production and trade across countries~\cite{Chen2011Gravity,Novy2012Gravity,Arvis2016} and performed analyses based on quantified trade costs.

Based on defined trade resistance $r_{i,j}$, the improved model used in this paper is depicted by the following formula:

\begin{equation}
{F_{i,j}} \propto \frac{{{{\left( {{m_i} \cdot {m_j}} \right)}^\alpha }}}{{{r_{i,j}}}}-\varepsilon_{ij}
\label{grav-r}
\end{equation}

where $m_i$ and $m_j$ are the gross domestic products of countries $i$ and $j$; $r_{i,j}$ is a defined composite variable; and $\alpha$ is the parameter to be estimated with the expectation maximization algorithm as the latent parameter in section \ref{EM}, $\varepsilon_{ij}$ is error term. Here, if we consider $r_{i,j}$ to be symmetric, the mechanism described in equation \ref{grav-r} is similar to Anderson's structural gravity model~\cite{anderson2002logit,anderson2004} but with a simpler expression. Here, $r_{i,j}$ is representative of trade resistance, which we use as a composite of all the other factors that affect trade volumes other than countries' GDP. Equation \ref{grav-r} indicates that the trade amount $F_{i,j}$ is proportional to $m_i$ and $m_j$ but inversely proportional to the integrated effective distance between them, denoted $r_{i,j}$. 

In contrast to the traditional gravity model, here, a country's geographical distance $d_{i,j}$ is replaced with trade resistance $r_{i,j}$. The new model not only captures proximity or distance in terms of geographical distance but also fully demonstrates the true and comprehensive relationships between entities in the system, which is significant for understanding the global economy, politics and culture~\cite{wang2018exploring}. 

In the literature, the trade cost measure can be derived from a broader range of models~\cite{anderson2002logit}, which have different methods and results in the parameter estimation, such as the elasticity of substitution $\sigma$~\cite{anderson2004}, the Frechet parameter $\vartheta$~\cite{Eaton2002Technology}, and the Pareto parameter $\gamma$~\cite{helpman2008estimating,Chaney2008Distorted,corcos2011}. With the estimated parameters and observed trade flow $F_{i,j}$, $m_i$ and $m_j$, the symmetrical trade resistance can be obtained from equation \ref{grav-r} using the least squares method.

However, the existing exogenous parameter estimation method will introduce unnecessary errors and doubts about validity. However, further analysis of trade resistance will inevitably involve the estimation of latent variables or parameters, and here, we use the EM algorithm from machine learning. In addition, there are many zero values in bilateral migration data, which is also a problem that has long puzzled researchers ~\cite{SilvanaThe2006,Santos2008,Ortega2013The,Fatima2019}. Here, we use the pseudo maximum likelihood (PML) method to preprocess the zero value flow; for details, see Appendix \ref{appendix: zero value}.

\subsection{Decomposing Trade Resistance through the Expectation Maximization Algorithm (EM)}\label{EM}

For each pair of countries $i$ and $j$, trade resistance $r_{i,j}$ is quantified by equation \ref{grav-r}, and we assume that trade resistance can be separated into two components. The data $\emph{\textbf{R}}=\{\ln r_{1,2},..., \ln r_{i,j},...\}$ can be divided into two categories: \uppercase\expandafter{\romannumeral1} is mainly related to natural factors such as geographical distance $d_{i,j}$, and \uppercase\expandafter{\romannumeral2} is affected more by artificial barriers than natural factors.

\begin{equation}
    \ln {r_{i,j}} = \left\{ \begin{array}{l}
a + b\ln {d_{i,j}} + {\eta _{i,j}}\quad \quad \quad \left( {r_{i,j}} \right) \in {\rm I}\\
{\xi _{i,j}}\quad \quad \quad \quad \quad \quad \quad \quad \quad\left( {r_{i,j}} \right) \in {\rm I}{\rm I}.
\end{array} \right.
\label{eeln}
\end{equation}
Here, $a,b$ are constants. $\eta _{i,j}$ and $\xi _{i,j}$ are normally distributed random variables with different means and standard deviations, $\eta _{i,j} \sim N(0,\sigma_1)$ and $\xi _{i,j} \sim N(\mu,\sigma_2)$. How should one estimate parameters $\Theta=\{\mu,\sigma_1,\sigma_2,a,b\}$ based on observed data $\emph{\textbf{R}}$ and place each $\ln r_{i,j}$ into the appropriate category?

To solve the parameter problem of two mixed distributions, we apply a commonly used method, namely, the EM algorithm. In statistics, the EM algorithm is an iterative method to find the maximum likelihood or maximum a posteriori (MAP) estimates of the parameters in statistical models, where the algorithm depends on unobserved latent variables~\cite{Dempster1977Maximum,Jordan1996On,Mclachlan2007The,TREVORHASTIE2008The}.

The EM algorithm seeks to obtain the MLE (maximum likelihood estimate) of the marginal likelihood by iteratively applying the expectation step (E step) and maximization likelihood step (M step), with $t=1,2,...$ representing the number of iterations. The detailed process is as follows:

\begin{itemize}
\item [\textbf{1.}] \textbf{Expectation step (E step):} In step $t$, based on the last estimation of the parameters $\hat{\Theta}^{(t-1)}$, calculate the expected value of the probability of belonging to a certain category.

Separately calculate the probabilities of observation $\ln r_{i,j}$ belonging to category \uppercase\expandafter{\romannumeral1} and category \uppercase\expandafter{\romannumeral2}. 
\begin{equation}
\begin{split}
&p_1(r_{i,j}\mid \hat{\Theta}^{(t-1)})=\frac{1}{{\sqrt {2\pi}\sigma_1}}\exp \frac{-[\ln r_{i,j}-(a+b\ln d_{i,j})]^2}{2\sigma_1^2},\\
&p_2(r_{i,j}\mid \hat{\Theta}^{(t-1)})=\frac{1}{{\sqrt {2\pi}\sigma_2}}\exp \frac{-[\ln r_{i,j}-\mu]^2}{2\sigma_2^2}.
\end{split}
\end{equation}

Then, normalize them as follows:

\begin{equation}
  \hat{\tau}_{i,j}^{(t)}=\frac{p_1(r_{i,j}\mid \hat{\Theta}^{(t-1)})}{p_1(r_{i,j}\mid \hat{\Theta}^{(t-1)})+p_2(r_{i,j}\mid \hat{\Theta}^{(t-1)})}.
  \end{equation}

The unobserved latent variables $\Theta_{\tau}=\{\tau_{1,2},\tau_{1,3},...,\tau_{i,j},...\}$, where $\tau_{i,j}$ ($0\le \tau_{i,j}\le 1$) represents the probability of trade resistance $\ln r_{i,j}$ belonging to category \uppercase\expandafter{\romannumeral1}.

\item [\textbf{2.}] \textbf{Maximization likelihood step (M step):} Based on the $\hat{\Theta}_{\tau}^{(t)}$ obtained from the E step, we find the parameter estimate $\Theta^{(t)}$ that maximizes this likelihood. The likelihood function $L$ of $\emph{\textbf{R}}$ occurring is multiplied by the expected probability of all trade resistances as follows:

\begin{equation}\nonumber
\begin{split}
L(\emph{\textbf{R}};\Theta,\Theta_{\tau}) =\prod \limits_{i\ne j}  \{\underbrace{\tau_{i,j} \cdot p_1({r_{i,j}\mid \Theta})}_{Category\quad \uppercase\expandafter{\romannumeral1}}+ \underbrace{(1-\tau_{i,j}) \cdot p_2({r_{i,j}\mid \Theta})}_{Category\quad \uppercase\expandafter{\romannumeral2}}\}.
\end{split}
\end{equation}

The optimum value of ${\Theta}^{(t)}$ based on $\emph{\textbf{R}}$ and $\hat{\Theta}_{\tau}^{(t)}$ can be calculated from that function:

\begin{equation}
  \begin{split}
   \hat{\Theta}^{(t)}&=\max_{\Theta}\log L(\emph{\textbf{R}};\Theta\mid\hat{\Theta}_{\tau}^{(t)})\\
   &=\max_{\Theta}\sum_{i\ne j} \log \{ \hat{\tau}_{i,j}^{(t)} \cdot p_1({r_{i,j}\mid \Theta})+ (1-\hat{\tau}_{i,j}^{(t)}) \cdot p_2({r_{i,j}\mid \Theta})\}.
   \end{split}
  \end{equation}

\end{itemize}

\subsection{Exploring Community Evolution based on the Extracted Backbone Trade Network}
\label{backbone network}

Here, we regard countries as the nodes, and the relationship between two nodes can be described by an edge. The reciprocal of trade resistance is the weight of the edge. Since trade resistance is symmetric for country pair $(i,j)$, the network is also symmetric. For node $i$, the node cluster coefficient $C_i$ is calculated by the equation below~\cite{HANSEN201131}:

\begin{equation}
\label{co}
    {C_i} = \frac{{2{e_{i}} }}{{{k_i}\left( {{k_i} - 1} \right)}}，
\end{equation}

where $e_{i}$ is the number of edges connected to adjacent nodes and $k_i$ denotes the number of nodes that are adjacent to node $i$. The cluster coefficient of the network is the mean of the cluster coefficients of all nodes.

To make the community classification more efficient, we apply the disparity filter method to obtain a backbone network~\cite{Serrano2009}.

\begin{equation}
    {\alpha _{ij}} = 1 - (k - 1)\int_0^{{p_{ij}}} {{{\left( {1 - x} \right)}^{k - 2}}dx < \alpha_s  }
\end{equation}

where ${\alpha _{ij}}$ is the probability of an edge between node $i$ and $j$, $k$ indicates the degree of a given node, ${p_{ij}}$ is the normalized weight of the edge and $\alpha_s$ is a significance level for the null hypothesis.

After extracting the backbone network, to classify the network into several communities, we apply the Louvain community detection algorithm~\cite{Blondel2008Fast} and evaluate the result using the Q index~\cite{Newman2004Detecting}.

\begin{equation}
    Q = \frac{1}{{2m}}\sum\limits_{i,j} {\left[ {{w_{i,j}} - \frac{{{A_i}{A_j}}}{{2m}}} \right]} \delta ({c_i},{c_j}),
\end{equation}

where $w_{ij}$ is the weight of the edge between nodes $i$ and $j$, ${A_i} = \sum\limits_j {{w_{i,j}}}$ is the sum of the weights of the edges attached to node $i$, $c_i$ is the community to which node $i$ belongs, and $\delta ({c_i},{c_j})$ is 1 if $c_i=c_j$ and 0 otherwise. $m = \frac{1}{2}\sum\nolimits_{i,j} {{w_{i,j}}}$ is the sum of the edge weights.
Based on the quantified trade resistances during the period 2007-2017, we can construct the backbone network of global trade for each year and attempt to explore the community classification of the network.

\section{Results and Discussion}

\subsection{Alienation of Global Trade Relationships}

\subsubsection{1. Trade Purity Indicator for Countries. }

\begin{figure}[hp]
    \centering
\subfigure[2007]{\includegraphics[width=.4\textwidth]{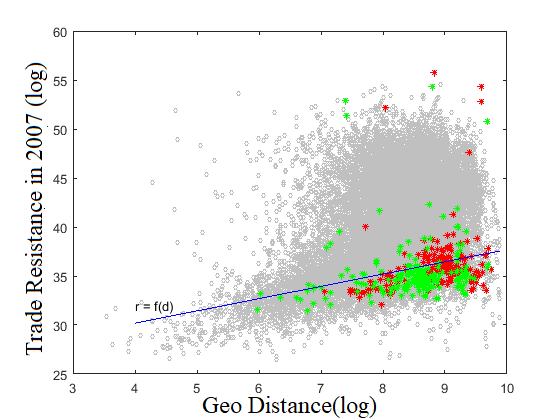}}
\subfigure[2007]{\includegraphics[width=.4\textwidth]{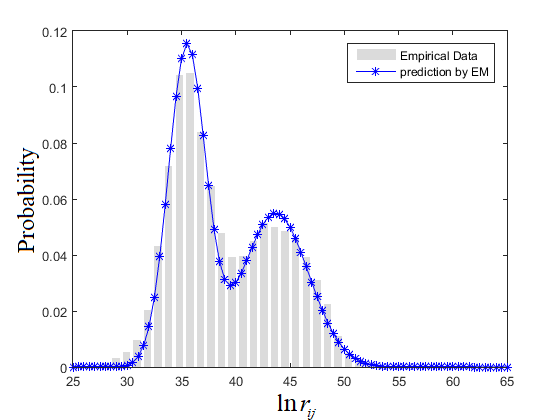}}
\subfigure[2012]{\includegraphics[width=.4\textwidth]{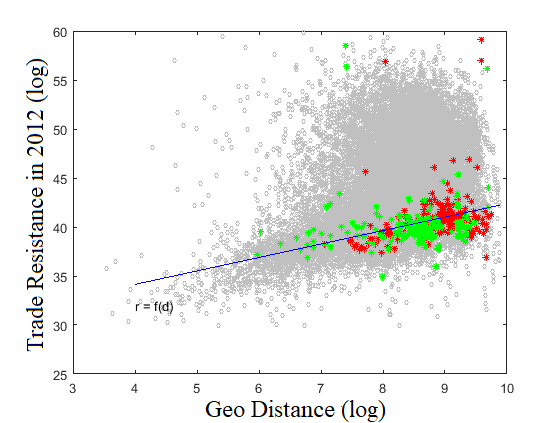}}
\subfigure[2012]{\includegraphics[width=.4\textwidth]{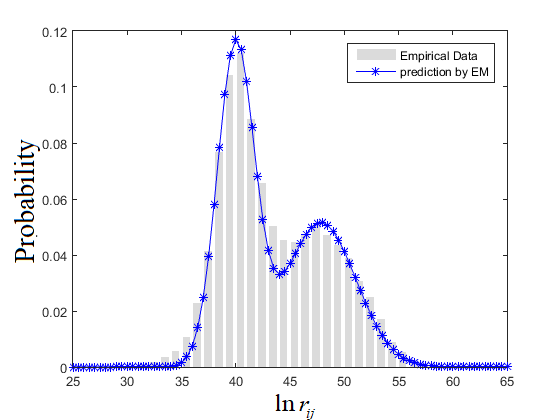}}
\subfigure[2017]{\includegraphics[width=.4\textwidth]{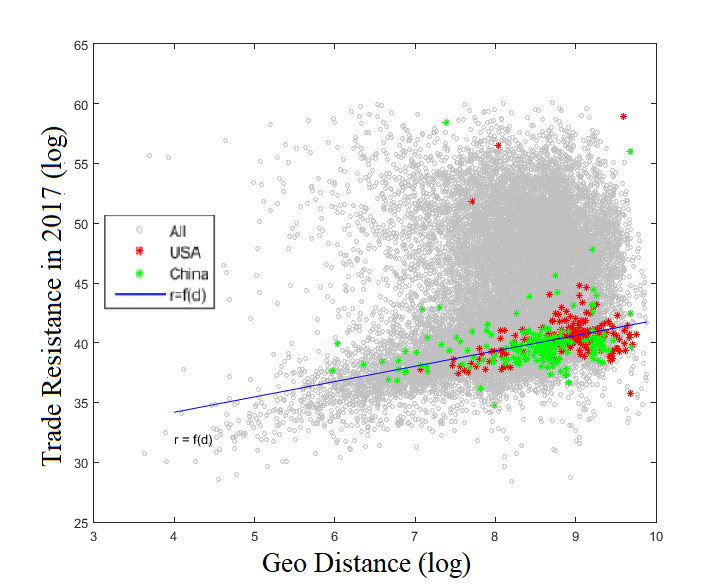}}
\subfigure[2017]{\includegraphics[width=.4\textwidth]{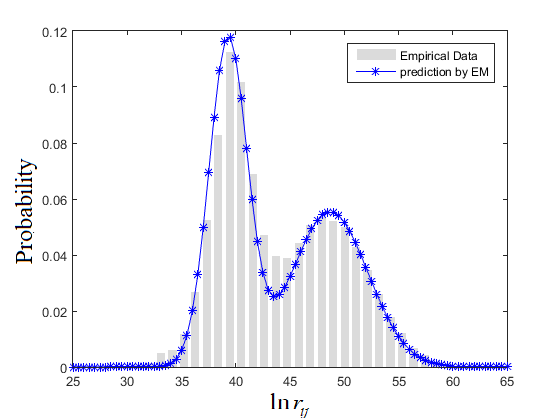}}
\caption{Fitting the Distribution Characteristics of Trade Resistance, based on the Hypothesis of Two Categories. (a, c, e) Gray dots show the trade resistance between countries around the world. (b, d, f) Gray bars express the trade resistance quantified from the extended gravity model; the blue dotted line is the fitted results with the EM algorithm.} 
    \label{fig:fitting}
\end{figure}

Based on the extended gravity model, we can quantify the international trade resistance $r_{i,j}$ for 198 entities (Figure \ref{fig:fitting}). We suppose that most trade resistance can be divided into two categories. The first has low expected barriers, which are mainly related to natural factors such as geographical distance, and the other includes countries with relatively high artificial trade barriers, such as trade restrictions, border blockades, cultural differences and political policies. 
It shows that most of the trade relations among the United States (red dot), China (green dot) and other countries belong to the first category, that is, most of the trade resistances are positively related to geographical distance, so they are concentrated near the blue dotted line (Figure \ref{fig:fitting} (a,c,e)). For the United States and China, only a small number of bilateral trade relations are affected by more artificial barriers.

Using the EM algorithm and the defined latent parameter $\theta=[a,b,\mu,\sigma_1,\sigma_2]$, we can fit the distribution of trade resistance $r_{i,j}$ well and obtain the characteristics of the two categories~\cite{Jordan1996On}. The fits of the distribution for 2007, 2012 and 2017 all pass the Kolmogorov-Smirnov test, and the parameters efficiently convert to the optimal values (Figure \ref{fig:fitting} (b,d,f)), which confirms our hypothesis of two categories of trade relations.

Here, the trade resistance of each pair has a probability of belonging to the limited trade resistance group (natural barriers, or category \uppercase\expandafter{\romannumeral1}). 
For each country $i$, we define the trade purity indicator $TPI_i$ by summing the probability that its trade relation $r_{i,j}$ belongs to category \uppercase\expandafter{\romannumeral1} as $TPI_i=1/N\sum_jP(z_{ij}=1\mid \hat{\theta})$, where $N$ is the number of countries, $z_{ij}$ equals 1 when the trade relations between $i$ and $j$ belong to category \uppercase\expandafter{\romannumeral1}, 0 otherwise. The TPI indicator provides a quantitative measure of the openness of a country's trade environment.

\subsubsection{2. Alienation of Trade Relationships between Countries. }

\begin{figure}[H]
    \centering
\subfigure[]{\includegraphics[width=.5\textwidth]{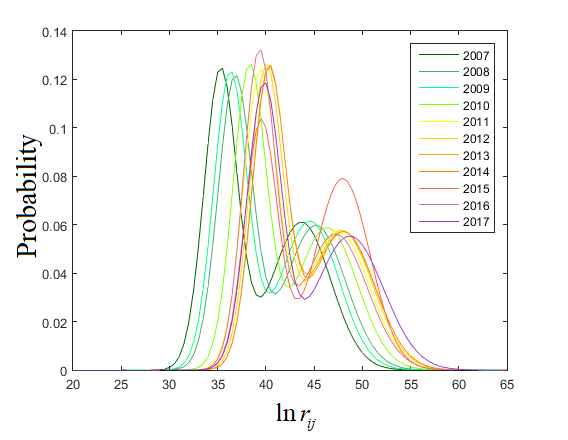}}
\subfigure[]{\includegraphics[width=.4\textwidth]{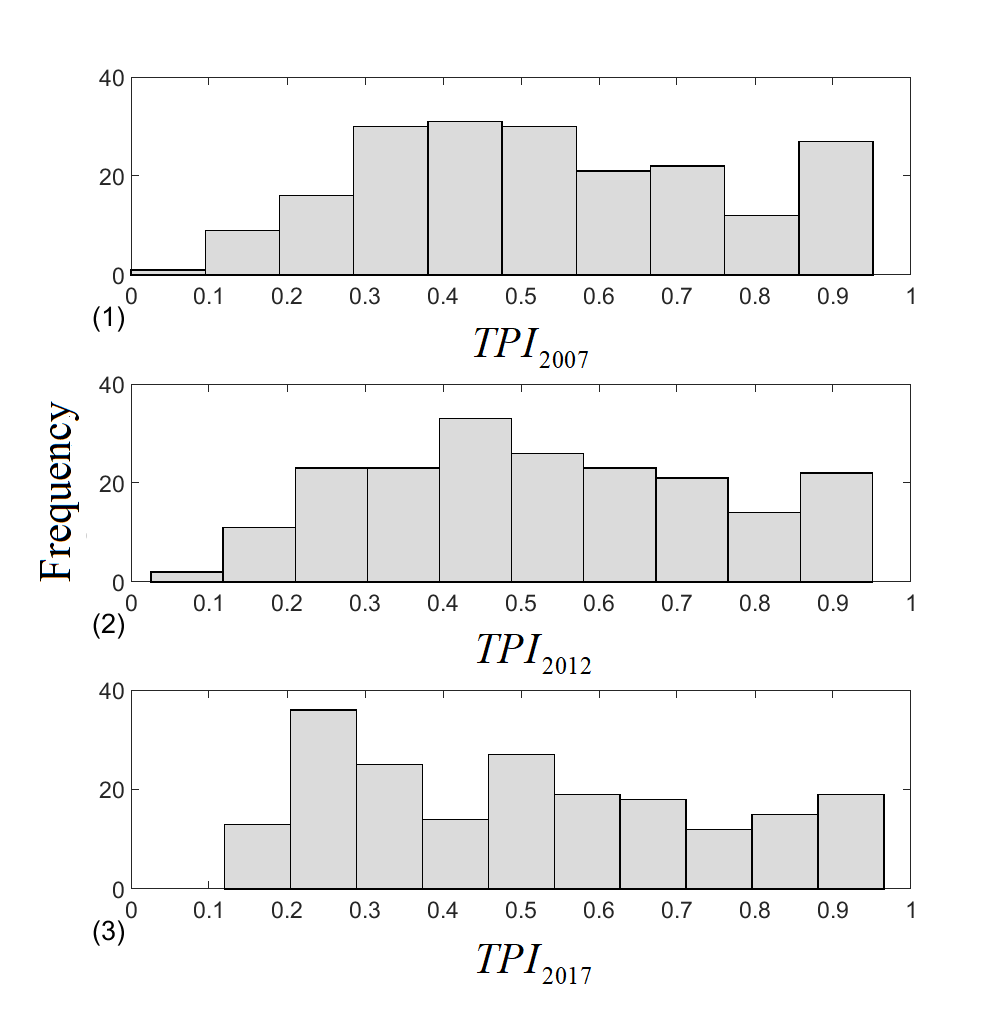}}
\caption{Evolution of Trade Resistance during the Period 2007-2017. (a) Distribution of trade resistance $\ln r_{ij}$. (b) Distribution of the trade purity indicator (TPI) in 2007, 2012, and 2017.}
    \label{fig:EVO}
\end{figure}

Figure \ref{fig:EVO} (a) shows the evolution of trade resistance. Different colors represent the distribution of trade resistance for corresponding years. With optimized parameters, in the decade considered, the distribution of trade resistance (the expectations of categories \uppercase\expandafter{\romannumeral1} and \uppercase\expandafter{\romannumeral2}) shifts to the right overall, which indicates an average increase in global trade resistance during the period 2007-2017.

Considering that global trade resistance could be affected by the growth of transportation costs or other factors, we also analyze the trend of the trade purity indicator from a more rigorous perspective. The distribution of the trade purity indicator (Figure \ref{fig:EVO}(b)) also indicates the alienation of the global trade network. Obviously, the mean TPI decreased in from 2007 to 2017.

The alienation of global trade is thought-provoking. In recent years, some scholars have highlighted this trend in international trade~\cite{Carden2008COMMERCE,MIDDLETON20071904}. To protect trade interests, some countries seek to maintain the friendly regional trade relations by signing trade agreements and creating trade unions. Since the 1990s, RTAs have proliferated, including regional unions with members that are geographically near one another (e.g., EU, NAFTA) and countries or regional blocs with diverse and geographically distant partners (e.g., ASEAN and BRI)~\cite{foster2011impact,Grossman2016}. The impact of RTAs has always interested politicians and scholars. Can RTAs adapt to such an international trade environment? Why might a government be willing to compromise its sovereignty and sign an agreement? The answer is interdependence. Based on the quantified TPI, we attempt to analyze the effects of regional trade unions in the following sections.

\subsection{Effects of Regional Trade Agreements (RTAs)}

The policies imposed by any government could affect the wellbeing not only of its own citizens but also those in other countries. Trade creation and trade diversion are common effects of RTAs identified in the recent literature~\cite{viner2014customs,panagariya2000preferential}, and in empirical work, their mixed effects are more complex; the results are difficult to quantify~\cite{burfisher2001impact,magee2008new}. This paper attempts to describe the effects of RTAs on both global and local trade relationships through a quantitative model and empirical analysis.

\subsubsection{1. Relatively Closer Trade Relationships between Union Members. }

Here, we analyze six typical RTAs, including those between the 28 EU countries, 52 OAU countries, 13 CARIFTA countries, 10 ASEAN countries, 3 NAFTA countries and 66 BRI countries.

First, we compare the trade resistance within and outside the six unions. In Table \ref{tab:average}, the average trade resistance between member countries is lower than that outside the unions. This demonstrates that the member countries of a union generally have closer trade relations with one another.

\begin{table}[tb]
	\centering 
	\caption{Average Trade Resistance Within and Outside Trade Unions}  
	\label{tab:average} 
\resizebox{\textwidth}{18mm}{
\begin{tabular}{c|c|c|c|c|c|c|c|c|c|c|c|c|c}
\hline\hline
\multirow{2}*{Year} & \multicolumn{2}{c|}{BRI} & \multicolumn{2}{c|}{EU} &\multicolumn{2}{c|}{OAU} & \multicolumn{2}{c|}{CARIFTA} & \multicolumn{2}{c|}{ASEAN} & \multicolumn{2}{c|}{NAFAT} & \multirow{2}*{World}\\ [0.5ex]
\cline{2-13}
 & Member & Others & Member & Others & Member & Others & Member & Others & Member & Others & Member & Others & \\\hline
2007 & 36.66 & 39.57 & 33.40 & 37.14 & 38.51 & 40.31 & 33.38 & 40.63 & 32.85 & 38.51 & 33.50 & 36.68 & 39.18 \\\hline
2008 & 38.16 & 41.10 & 34.92 & 38.59 & 40.01 & 41.80 & 34.59 & 41.97 & 34.57 & 39.87 & 35.09 & 38.17 & 40.66 \\\hline
2009 & 37.65 & 40.47 & 34.27 & 38.08 & 39.55 & 41.16 & 34.33 & 41.42 & 33.83 & 39.27 & 34.34 & 37.45 & 40.06 \\\hline
2010 & 39.45 & 42.33 & 36.30 & 40.04 & 41.24 & 42.98 & 36.42 & 43.43 & 35.95 & 41.10 & 36.71 & 39.71 & 41.97 \\\hline
2011 & 40.68 & 43.47 & 37.37 & 41.13 & 42.43 & 44.11 & 37.18 & 44.45 & 37.01 & 42.31 & 37.92 & 40.84 & 43.07 \\\hline
2012 & 41.22 & 43.93 & 37.86 & 41.61 & 42.83 & 44.62 & 37.63 & 44.90 & 37.61 & 42.63 & 38.54 & 41.31 & 43.54 \\\hline
2013 & 41.26 & 44.00 & 37.85 & 41.68 & 43.02 & 44.74 & 37.79 & 45.07 & 37.74 & 42.54 & 38.52 & 41.39 & 43.66 \\\hline
2014 & 41.27 & 43.94 & 37.81 & 41.44 & 43.38 & 44.73 & 37.95 & 45.11 & 37.41 & 42.71 & 38.48 & 41.39 & 43.65 \\\hline
2015 & 41.51 & 44.21 & 36.91 & 42.03 & 44.69 & 45.37 & 39.87 & 45.34 & 36.85 & 42.93 & 37.58 & 42.32 & 44.13 \\\hline
2016 & 40.18 & 42.96 & 36.79 & 40.45 & 42.99 & 43.95 & 38.16 & 44.46 & 36.49 & 41.65 & 37.41 & 40.33 & 42.79 \\\hline
2017 & 41.59 & 44.23 & 37.27 & 41.41 & 43.90 & 45.14 & 38.75 & 45.47 & 39.47 & 43.44 & 37.89 & 40.80 & 43.98 \\\hline

\end{tabular}}
\end{table}

\begin{figure}[h]
    \centering
    \includegraphics[width=0.8\textwidth]{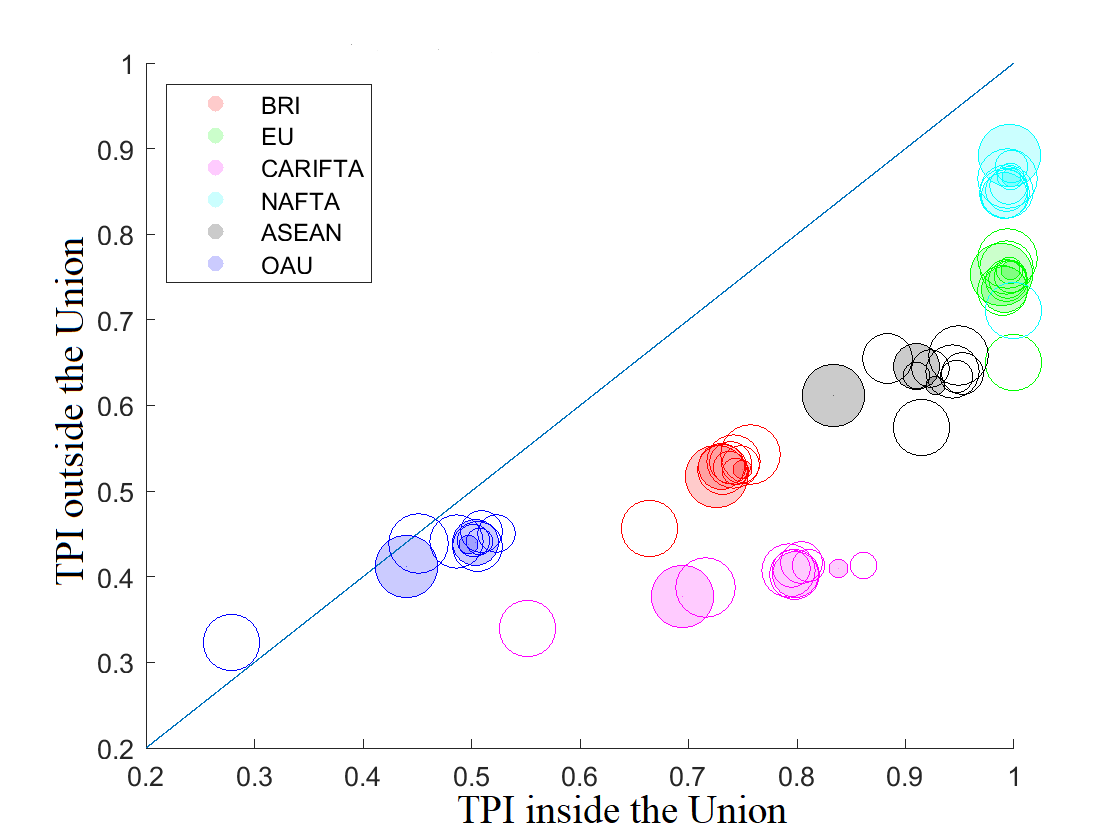}
\caption{Trade Purity Indicator Within and Outside Unions During the Period 2007-2017. The X-coordinate expresses the TPI inside the union, and the y-coordinate expresses the TPI outside the union.}
    \label{fig:meanTPI}
\end{figure}

In addition, in Figure \ref{fig:meanTPI}, with the x-coordinate expressing the TPI within the union and the y-coordinate expressing the TPI beyond the union, the size of the dots is proportional to its proximity to the present (TPI in 2007 has the smallest radius, and TPI in 2017 has the largest radius); the solid three spots of the same color indicate the TPI in 2007, 2012 and 2017. Obviously, most spots are located below the diagonal, which means that the relationships between union members are closer than those with other countries outside the union.

Therefore, it indicates that all trade unions help to lower average trade resistances and create closer trade relations among the members compared with other countries.

\subsubsection{2. Decreasing Trend in Trade Relationships for Union Members. }

These six unions can be divided into two types. Specifically, the EU and NAFTA are type one, and most countries in these unions are developed countries. For these two RTAs, the spots move vertically over time (Figure \ref{fig:meanTPI}). The TPI inside the unions barely changes, but the TPI outside the unions fluctuates and tended to increase. BRI, OAU, CARIFTA and OAU are type two, and the spots of these unions move towards the bottom left. In brief, by comparing the TPIs in 2007 and 2017, except for the EU and NAFTA, the TPIs within unions all declined. The trade environments of the EU and NAFTA are more friendly than those in the other four unions.

In Figure \ref{fig:TR-6} (in the Appendix), we can more clearly see these two types of unions. The red labels indicate a trade deficit, while blue labels indicate a trade surplus, and the size of spots represents the net trade flow (exports minus imports). The EU and NAFTA (Figure \ref{fig:TR-6}(a)(e)) have fewer member countries, and have higher economic development and surplus trade flows. Therefore, the dots are highly concentrated. 
Other unions (BRI, OAU, CARIFTA and ASEAN (Figure \ref{fig:TR-6}(b)(c)(d)(f)) are more uneven, as the dots distributed from low TPI to high TPI, and some member countries have trade surplus, while the others have a trade deficit. 
In addition, this indicates that the countries with surplus trade flows (blue label) have a higher TPI both inside and outside their unions.

\subsection{Comparison of Trade Unions and Trade Communities}

Trade unions are formed through agreements signed by countries. With the development of globalization, it is worth further exploring whether they can reflect real trade affinity. As mentioned in section \ref{backbone network}, we extract the backbone of the global trade network based on quantified trade resistance and classify it into several communities. Trade communities are obtained from the analysis of the network structure, which can objectively describe the trade relationships between countries.

\subsubsection{1. Communities in the Global Trade Network. }

\begin{figure}[h]
    \centering
\subfigure[2007]{\includegraphics[width=.8\textwidth]{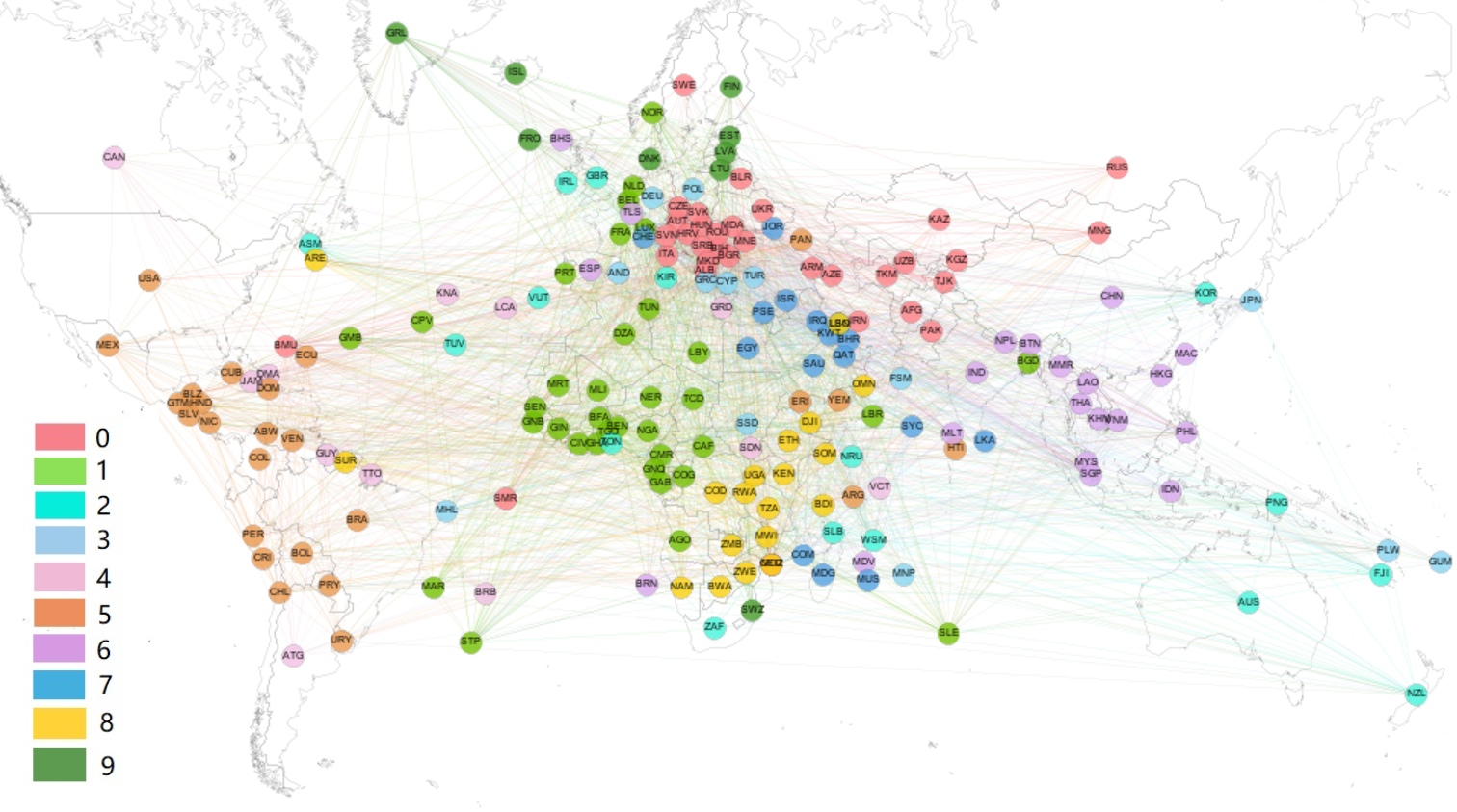}}
\subfigure[2017]{\includegraphics[width=.8\textwidth]{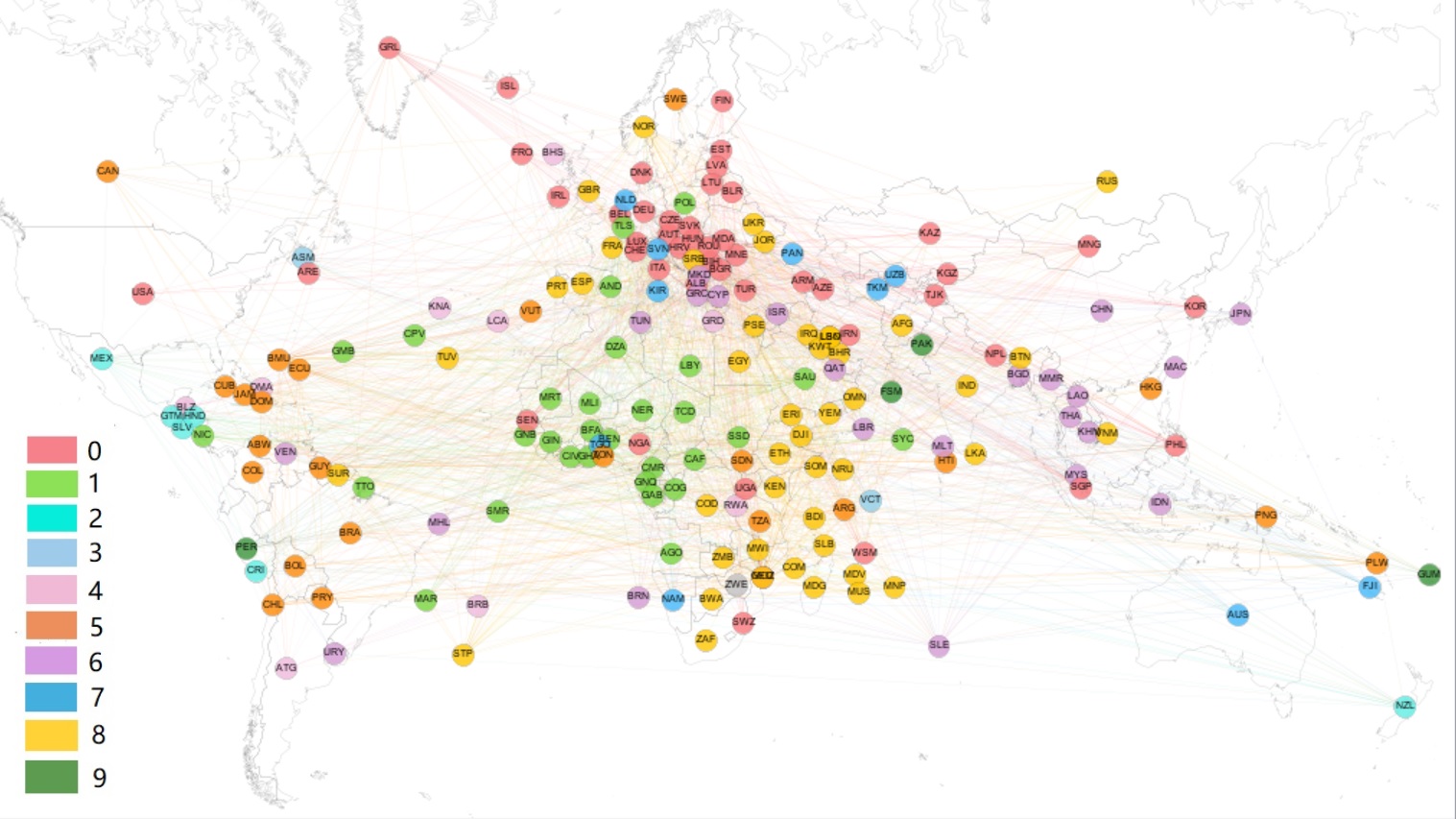}}
\caption{Communities in the Global Trade Network in 2007 (a) and 2017 (b).}
    \label{fig:community20072017}
\end{figure}

In Figure \ref{fig:community20072017}, the nodes that share the same color are assigned to the same community. The modularity of classification is $Q=0.780$ in 2007 and $Q=0.769$ in 2017, which means that the classification is credible. There were some structural changes between 2007 and 2017.

First, in 2007, the communities show significant regionality. The map (Figure \ref{fig:community20072017} (a)) shows that countries on the same continent are more likely to be clustered in the same community, confirming that geographical characteristics play an important role in forming trade patterns. For most members, the six trade unions are signed among regional countries, and based on their relatively close trade relations, it is not difficult to understand that most members of RTAs are clustered in the same community. In 2017, the distribution of community members was more divergent. With the development of globalization, trade between countries is no longer restricted by geographical or transportation factors.

Second, from the empirical results, over these ten years, the network density decreased, which means that countries in the global trade network are connected more loosely (the cluster coefficient changed from 0.1370 to 0.1006).

There is another interesting phenomenon. Some countries are on the same continent and belong to the same RTA but are more closely related to countries in other unions than the members of their RTAs. Most African countries have multiple RTA memberships~\cite{Gupta2007Regional}, and the continent's east and west coasts belong to different marine routes in the global marine transport network~\cite{Zhang2018Statistical}. Therefore, it is easy to understand why eastern and western Africa are clustered into different communities. France, Spain, Portugal, and Belgium are EU countries, but they are classified into the community where most members are African countries. This shows that they have closer trade relations with African countries than with other EU members, which may be due to language, culture, colonial influence and their trade structures.

Here, we apply the external-internal index (E-I Index) and compare regional trade cohesion and global trade cohesion as follows:
\begin{equation}
\begin{split}
   \text{E-I index}_{(degree)}&=-\frac{EK-IK}{EK+IK} \\
   \text{E-I index}_{(weight)}&=-\frac{EW-IW}{EW+IW} 
\end{split}
\end{equation}

External edges connect nodes from different communities, while internal edges connect two nodes belonging to the same community. $EK$ and $IK$ are the sum of external and internal degrees for all nodes; $EW$ and $IW$ are the sum of external and internal weights for all nodes. Based on the results of the backbone network in 2007 and 2017, the E-I index (degree) dropped from 0.2711 to 0.1000, and the E-I index (weight) increased from -0.1019 to 0.0281. The relationships in the global trade network are more diversified, but trade intensity is concentrated in local communities.

\subsubsection{2. Correlation and Evolution of Unions and Communities. }

\begin{figure}[hp]
    \centering
\subfigure[2007]{\includegraphics[width=.8\textwidth]{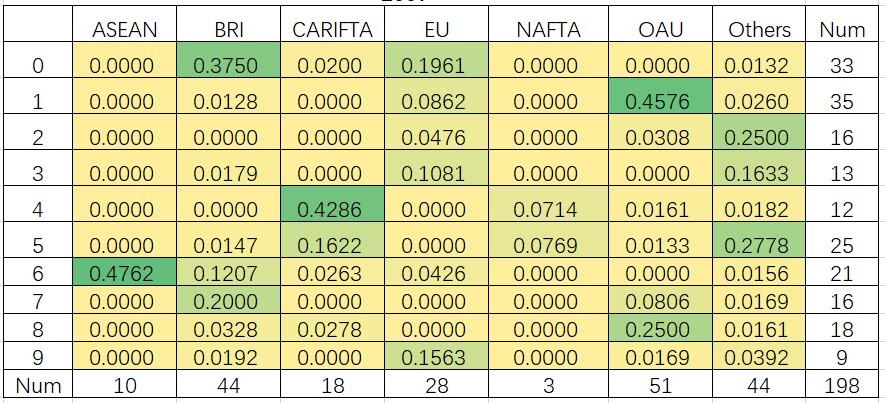}}
\subfigure[2017]{\includegraphics[width=.8\textwidth]{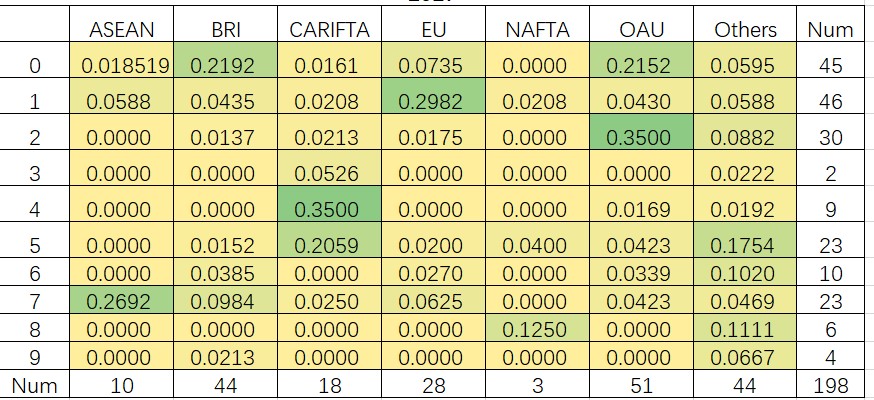}}
\caption{The Similarity Matrix between Trade Unions and Trade Communities in 2007 (a) and 2017 (b). `Num' means the number of members in the corresponding trade union or community.}
    \label{fig:EI20072017}
\end{figure}

We have identified the trade unions resulting from negotiations between countries, and the trade communities clustered from the empirical data. What is the correlation between them? Do the members of trade unions truly have closer trade relations? Which trade unions have no obvious effect on restraining and helping member countries? To answer these questions, we measure the correlation coefficient between the members of trade unions and trade communities. Figure \ref{fig:EI20072017} shows the matrix of the Jaccard similarity coefficient of six trade unions and ten typical communities. `Others' indicates countries that do not belong to the six unions. Green color means a greater correlation and a higher commonality of members between trade unions and communities. In contrast, yellow color means that the members of the union and community are basically different.

In general, the similarity matrices of 2007 and 2017 have similar structures. Each trade union has only one or two grids with a great deal of green, which indicates that some trade unions and communities have high consistency in membership. Several very green grids are shown in Figure \ref{fig:EI20072017} (a), which presents the overlap of ASEAN and community 6, BRI and community 0, CARIFTA and community 4, OAU and community 1, etc. The EU and NAFTA are relatively `free' trade unions, and their members are not limited to one or two communities, which overlap with several separate communities. In addition, in 2007, the `Other' countries that do not belong to six trade unions are also relatively concentrated in three communities, with a certain aggregation, and it is quite different from the results from 2017.

The similarity was higher in 2007; that is, the trade unions were more similar to the actual trading clustering result. In 2017, the role of the trade unions weakened. For most trade unions, the maximum matching of members to communities is decreasing. Here, the EU and NAFTA remain exceptions, having relatively higher similarity with communities 1 and 8. This might be due to their mature trading background. In addition, the TPIs of the EU and NAFTA remained stable, while the TPIs of other trade unions decreased (Figure \ref{fig:meanTPI}). TPI indicates a trade-friendly relationship with other countries, while communities also reflect the different trade relations between countries inside and outside the community. Therefore, it is reasonable and scientific to conclude that EU and NAFTA have particularities in both results. Compared with 2007, the community structure of ``Other'' countries has become more decentralized.

In short, RTAs appear to have an impact that strengthened the formation of true trade relations~\cite{Reyes2014}. Based on the similarity matrix, each trade union is mainly concentrated in one or two communities. However, in 2017, this kind of consistency clearly weakened, and multilateral trade liberalization has accelerated over the past decade.

\section{Conclusion and Discussion}

The innovation of this paper is to study and describe the trade purity relationships between countries, considering some other typical factors, such as economic volume, geographical distance, overall increased transportation and labor costs, are separated. In addition, this paper does not use the exogenous parameter estimation method, and we define latent parameters and use the likelihood function and EM algorithm to quantify and analyze the trade purity indicator more scientifically and effectively. In brief, the extended model prompts the development of the gravity model in theoretical research on international trade.

In the empirical analysis, some unobserved characteristics of the trade relationships can simultaneously be defined and optimized, and the analysis uses a trade purity indicator to describe the trade environments and positions of countries in both regional and global trade relationships.

With the data from the UN Comtrade Database, we quantify the international trade resistance of 198 countries/districts. This analysis shows that the trade relationships of the 198 entities can be divided into two categories. The trade resistance of countries in category \uppercase\expandafter{\romannumeral1} has an approximate log-linear relation with geographical distance, and these countries have a relatively open and friendly trade environment, where the main trade barriers are natural factors. The countries in category \uppercase\expandafter{\romannumeral2} have higher artificial trade barriers, and countries with poor trading environments frequently fall into category \uppercase\expandafter{\romannumeral2}. Here, we obtain well fitted results using the EM algorithm from machine learning. All latent variables converge rapidly to optimal points, which validates the extended gravity model proposed in this paper.

In addition, this paper defines and identifies a trade purity indicator for different RTA countries during the period 2007-2017. It can describe the true trade environment and relationships. Countries with higher indicators have friendly trade environments and obtain large trade flows, such as the United States, China, Japan, Korea, South Africa, Singapore, Australia, and Malaysia. For these countries, most trade partnerships are mainly related to natural factors such as geographical distance, and they have no obvious trade barriers. The analysis of the indicator and its evolution could help to research the characteristics and trends of international trade. This indicates that although the global and most regional trade relations gradually deteriorated over the period 2007-2017, the RTAs bring closer trade relations between members, especially contributing to the relative prosperity of the EU and NAFTA.

Finally, based on the trade resistance matrix, we build a network mapping the relationship of 198 countries/districts. The Louvain community detection method identifies several communities in the global trade network. Here, we analyze the effects of RTAs by comparing the members of trade unions and communities. The results show that the representative RTAs constitute the core structure of international trade network, but the role of trade unions has weakened and multilateral trade liberalization has accelerated in the past decade. This means that more countries have recently tended to expand their trading partners outside their unions rather than limit their trading activities to their RTAs.

\clearpage

\section*{Conflicts of Interest}
The authors declare no conflict of interest including any financial, personal or other relationships with other people or organizations.

\section*{Funding Statement}
This work was supported by the Chinese National Natural Science Foundation (71701018, 61673070), the National Social Sciences Fund, China (14BSH024), and the Beijing Normal University Cross-Discipline Project.

\section*{Data Availability statement}
 
The empirical data used in this paper could be downloaded from sources listed in Table \ref{table1}. Or please contact with Dr. Xiaomeng Li (lixiaomeng@bnu.edu.cn) to request the data.

\section*{Acknowledgement}
We appreciate comments and helpful suggestions from Prof. Zengru Di, Honggang Li, Handong Li and Jiang Zhang.

\clearpage

\bibliographystyle{splncs04}
\bibliography{main}

\begin{thebibliography}{10}
\providecommand{\url}[1]{\texttt{#1}}
\providecommand{\urlprefix}{URL }
\providecommand{\doi}[1]{https://doi.org/#1}

\bibitem{anderson2011gravity}
Anderson, J.E.: The gravity model. Annu. Rev. Econ.  \textbf{3}(1),  133--160
  (2011)

\bibitem{Anderson2010The}
Anderson, J.E., Yotov, Y.V.: The changing incidence of geography. American
  Economic Review  \textbf{100}(5),  2157--2186 (2010)

\bibitem{anderson2002logit}
Anderson, S.P., Goeree, J.K., Holt, C.A.: The logit equilibrium: A perspective
  on intuitive behavioral anomalies. Southern Economic Journal  \textbf{69}(1),
   21--47 (2002)

\bibitem{Anderson2003}
Anderson, James~E, v.W.: Gravity with gravitas : A solution to the border
  puzzle 1. American Economic Review  \textbf{93}(1),  170--92 (2003)

\bibitem{anderson2004}
Anderson J~E, V.W.E.: Trade costs. Journal of Economic Literature
  \textbf{42}(3),  691--751 (2004)

\bibitem{Arvis2016}
Arvis J~F, Duval~Y, S.B.e.a.: Trade costs in the developing world: 1996–2010.
  World Trade Review  \textbf{15}(03),  451--474 (2016)

\bibitem{Grossman2016}
Bagwell, K., Staiger, R.W.: Chapter 7 - the purpose of trade agreements. In:
  Handbook of Commercial Policy, Handbook of Commercial Policy, vol.~1, pp. 379
  -- 434. North-Holland (2016)

\bibitem{Bao2019The}
Bao, X., Wang, X.: The evolution and reshaping of globalization: A perspective
  based on the development of regional trade agreements. China \& World Economy
   \textbf{27}(1),  51--71 (2019)

\bibitem{bhagwati1996the}
Bhagwati, J.N., Panagariya, A.: The theory of preferential trade agreements:
  Historical evolution and current trends. The American Economic Review
  \textbf{86}(2),  82--87 (1996)

\bibitem{Blondel2008Fast}
Blondel, V., Guillaume, J.L., Lambiotte, R., Lefebvre, E.: Fast unfolding of
  communities in large networks. Journal of Statistical Mechanics Theory and
  Experiment  \textbf{2008} (04 2008)

\bibitem{burfisher2001impact}
Burfisher, M.E., Robinson, S., Thierfelder, K.: The impact of nafta on the
  united states. Journal of Economic perspectives  \textbf{15}(1),  125--144
  (2001)

\bibitem{burger2009specification}
Burger, M., Van~Oort, F., Linders, G.J.: On the specification of the gravity
  model of trade: zeros, excess zeros and zero-inflated estimation. Spatial
  Economic Analysis  \textbf{4}(2),  167--190 (2009)

\bibitem{Carden2008COMMERCE}
Carden, A., W.: Commerce and culture in the global economy. Journal of
  Interdisciplinary Studies  \textbf{20}(1),  21--36 (2008)

\bibitem{Carr2006Revisiting}
Carrère, C.: Revisiting the effects of regional trade agreements on trade
  flows with proper specification of the gravity model. European Economic
  Review  \textbf{50}(2),  223--247 (2006)

\bibitem{Chaney2008Distorted}
Chaney, T.: Distorted gravity: The intensive and extensive margins of
  international trade. American Economic Review  \textbf{98}(4),  1707--1721
  (2008)

\bibitem{Chen2011Gravity}
Chen, N., Novy, D.: Gravity, trade integration, and heterogeneity across
  industries. Journal of International Economics  \textbf{85}(2),  206--221
  (2011)

\bibitem{Cipollina2010Reciprocal}
Cipollina, M., Salvatici, L.: Reciprocal trade agreements in gravity models: A
  meta-analysis. Review of International Economics  \textbf{18}(1),  63--80
  (2010)

\bibitem{corcos2011}
Corcos~G, Del Gatto~M, M.G.e.a.: Productivity and firm selection: Quantifying
  the 'new' gains from trade. SSRN Electronic Journal  \textbf{122}(561),
  754--798 (2011)

\bibitem{Dai2014}
Dai, M., Yotov, Y.V., Zylkin, T.: {On the trade-diversion effects of free trade
  agreements}. Economics Letters  \textbf{122}(2),  321--325 (2014)

\bibitem{Deardorff2004Local}
Deardorff, A.V.: Local comparative advantage: Trade costs and the pattern of
  trade. Working Papers  \textbf{10}(1),  9–35 (2004)

\bibitem{Dempster1977Maximum}
Dempster, A.P., Laird, N.M., Rubin, D.B.: Maximum likelihood from incomplete
  data via the em algorithm. Journal of the Royal Statistical Society
  \textbf{39}(1),  1--38 (1977)

\bibitem{Eaton2002Technology}
Eaton, J., Kortum, S.: Technology, geography, and trade. Econometrica
  \textbf{70}(5),  1741--1779 (2002)

\bibitem{viner2014customs}
Emmett, Ross, B.: Jacob viner, the customs union issue. Journal of the History
  of Economic Thought  \textbf{37}(04),  629--630 (2015)

\bibitem{Fatima2019}
Fatima Olanike~Kareem, O.I.K.: The issues of zero values in trade data and
  modelling. Macro Management \& Public Policies  \textbf{01}(01),  36--50
  (2019)

\bibitem{Fatima2019zeroes}
Fatima Olanike~Kareem, O.I.K.: The issues of zero values in trade data and
  modelling. Macro Management \& Public Policies  \textbf{01}(01),  36--50
  (2019)

\bibitem{flowerdew1982method}
Flowerdew, R., Aitkin, M.: A method of fitting the gravity model based on the
  poisson distribution. Journal of regional science  \textbf{22}(2),  191--202
  (1982)

\bibitem{foster2011impact}
Foster, N., Poeschl, J., Stehrer, R.: The impact of preferential trade
  agreements on the margins of international trade. Economic Systems
  \textbf{35}(1),  84--97 (2011)

\bibitem{Ortega2013The}
Francesc, O., Giovanni, P.: The effect of income and immigration policies on
  international migration. Migration Studies  \textbf{1}(1), ~1 (2013)

\bibitem{Ghosh2004Are}
Ghosh, S., Yamarik, S.: Are regional trading arrangements trade creating?:an
  application of extreme bounds analysis. Journal of International Economics
  \textbf{63}(2),  369--395 (2004)

\bibitem{Gupta2007Regional}
Gupta, S., Yang, Y.: Regional trade agreements in africa: Past performance and
  way forward. African Development Review  \textbf{19}(3),  399–431 (2007)

\bibitem{HANSEN201131}
Hansen, D.L., Shneiderman, B., Smith, M.A.: Chapter 3 - social network
  analysis: Measuring, mapping, and modeling collections of connections. In:
  Analyzing Social Media Networks with NodeXL, pp. 31--50. Morgan Kaufmann
  (2011)

\bibitem{TREVORHASTIE2008The}
Hastie, T., Tibshirani, R., Friedman, J.H.: The elements of statistical
  learning: data mining, inference, and prediction, vol.~27. Springer (2005)

\bibitem{Head2014Chapter}
Head, K., Mayer, T.: Chapter 3 – gravity equations: Workhorse,toolkit, and
  cookbook. In: Handbook of International Economics, vol.~4, pp. 131--195.
  Elsevier (2014)

\bibitem{helpman2008estimating}
Helpman, E., Melitz, M., Rubinstein, Y.: Estimating trade flows: Trading
  partners and trading volumes. The quarterly journal of economics
  \textbf{123}(2),  441--487 (2008)

\bibitem{RePEc:ags:eprcrs:166082}
Isaac, S., Lawrence, O.: Trade creation and diversion effects of the east
  african community regional trade agreement: A gravity model analysis.
  Research Series 166082, Economic Policy Research Centre (EPRC) (2013)

\bibitem{isard1954location}
Isard, W.: Location theory and trade theory: Short-run analysis. The Quarterly
  Journal of Economics  \textbf{68}(2),  305--320 (1954)

\bibitem{SilvanaThe2006}
J., M., C., Santos, Silva, Silvana, Tenreyro: The log of gravity. Review of
  Economics \& Statistics  \textbf{88}(4),  641--658 (2006)

\bibitem{Jordan1996On}
Jordan, M.I., Lei, X.: On convergence properties of the em algorithm for
  gaussian mixtures. Neural Computation  \textbf{8}(1),  129--151 (1996)

\bibitem{jung2008gravity}
Jung, W.S., Wang, F., Stanley, H.E.: Gravity model in the korean highway. EPL
  (Europhysics Letters)  \textbf{81}(4),  48005 (2008)

\bibitem{Krisztin2015}
Krisztin, Tamás, F.M.M.: The gravity model for international trade:
  Specification and estimation issues. Spatial Economic Analysis
  \textbf{10}(4),  451--470 (2015)

\bibitem{Liu2018}
Liu, A., Lu, C., Wang, Z.: The roles of cultural and institutional distance in
  international trade: Evidence from china's trade with the belt and road
  countries. China Economic Review pp. 1--17 (2018)

\bibitem{magee2008new}
Magee, C.S.: New measures of trade creation and trade diversion. Journal of
  International Economics  \textbf{75}(2),  349--362 (2008)

\bibitem{manning2001estimating}
Manning, W.G., Mullahy, J.: Estimating log models: to transform or not to
  transform? Journal of health economics  \textbf{20}(4),  461--494 (2001)

\bibitem{martinez2011log}
Mart{\'\i}nez-Zarzoso, I.: The log of gravity revisited. Applied Economics
  \textbf{45}(3),  311--327 (2011)

\bibitem{Mclachlan2007The}
Mclachlan, G.J., Krishnan, T.: The EM Algorithm and Extensions, Second Edition.
  John Wiley \& Sons, Inc. (2007)

\bibitem{MIDDLETON20071904}
Middleton, A.: Globalization, free trade, and the social impact of the decline
  of informal production: The case of artisans in quito, ecuador. World
  Development  \textbf{3}(11),  1904 -- 1928 (2007)

\bibitem{Newman2004Detecting}
Newman, M.E.J.: Detecting community structure in networks. European Physical
  Journal B  \textbf{38}(2),  321--330 (2004)

\bibitem{Novy2012Gravity}
Novy, D.: Gravity redux: Measuring international trade costs with panel data.
  Economic Inquiry  \textbf{51}(1),  101--121 (2012)

\bibitem{panagariya2000preferential}
Panagariya, A.: Preferential trade liberalization: the traditional theory and
  new developments. Journal of Economic literature  \textbf{38}(2),  287--331
  (2000)

\bibitem{Reyes2014}
Reyes, J., Wooster, R., Shirrell, S.: Regional trade agreements and the pattern
  of trade: A networks approach. World Economy  \textbf{37}(8),  1128--1151
  (2014)

\bibitem{Santos2008}
Santos~Silva, J. M.~C., .T.S.: Trading partners and trading volumes:
  implementing the helpman-melitz-rubinstein model empirically. Oxford Bulletin
  of Economics \& Statistics  \textbf{77}(1),  93--105 (2008)

\bibitem{Serrano2009}
Serrano, M.{\'{A}}., Bogu{\~{n}}{\'{a}}, M., Vespignani, A.: Extracting the
  multiscale backbone of complex weighted networks. Proceedings of the National
  Academy of Sciences of the United States of America  \textbf{106}(16),
  6483--6488 (2009)

\bibitem{Silva2006The}
Silva, J.M.C.S., Tenreyro, S.: The log of gravity. Review of Economics \&
  Statistics  \textbf{88}(4),  641--658 (2006)

\bibitem{Song2017Topological}
Song, Z., Che, S., Yang, Y.: Topological relationship between trade network in
  the belt and road initiative area and global trade network. Progress in
  Geography  \textbf{36}(11),  1340--1348 (2017)

\bibitem{Tadesse2017Does}
Tadesse, B., White, R., Huang, Z.: Does china's trade defy cultural barriers?
  International Review of Applied Economics  \textbf{31}(3),  398--428 (2017)

\bibitem{tinbergen1963shaping}
Tinbergen, J.: Shaping the world economy. The International Executive
  \textbf{5}(1),  27--30 (1963)

\bibitem{Vicard2009On}
Vicard, V.: On trade creation and regional trade agreements: Does depth matter?
  Review of World Economics  \textbf{145}(2),  167--187 (2009)

\bibitem{wang2018exploring}
Wang, K., Li, X., Wang, X., Chen, Q., Bao, J.: Exploring the true relationship
  among countries from flow data of international trade and migration.
  International Conference on Complex Systems pp. 476--485 (2018)

\bibitem{Yang2014}
Yang, S., Martinez-Zarzoso, I.: A panel data analysis of trade creation and
  trade diversion effects: The case of asean–china free trade area. China
  Economic Review  \textbf{29}(C),  138--151 (2014)

\bibitem{Zhang2019}
Zhang, C., Fu, J., Pu, Z.: A study of the petroleum trade network of countries
  along ``the belt and road initiativ''. Journal of Cleaner Production
  \textbf{222},  593--605 (2019)

\bibitem{Zhang2018Statistical}
Zhang, W., Deng, W., Wei, L.: Statistical properties of links of network: A
  survey on the shipping lines of worldwide marine transport network. Physica A
  Statistical Mechanics \& Its Applications  \textbf{502},  218--227 (2018)

\bibitem{Zhang2006Network}
Zhang, Y.Q., Hai, Z., Zhang, W.B., Tie-Jun, H.A.: Network characteristics of
  urban population migration. Journal of Northeastern University
  \textbf{27}(2),  169--172 (2006)

\bibitem{Zhong2014The}
Zhong, W., An, H., Gao, X., Sun, X.: The evolution of communities in the
  international oil trade network. Physica A Statistical Mechanics \& Its
  Applications  \textbf{413}(11),  42--52 (2014)

\end{thebibliography}

\section*{Appendix}
\appendix
\renewcommand{\appendixname}{Appendix~\Alph{section}}

\section{Figures}

\begin{figure}[H]
    \centering
    \subfigure[EU]{\includegraphics[width=.4\textwidth]{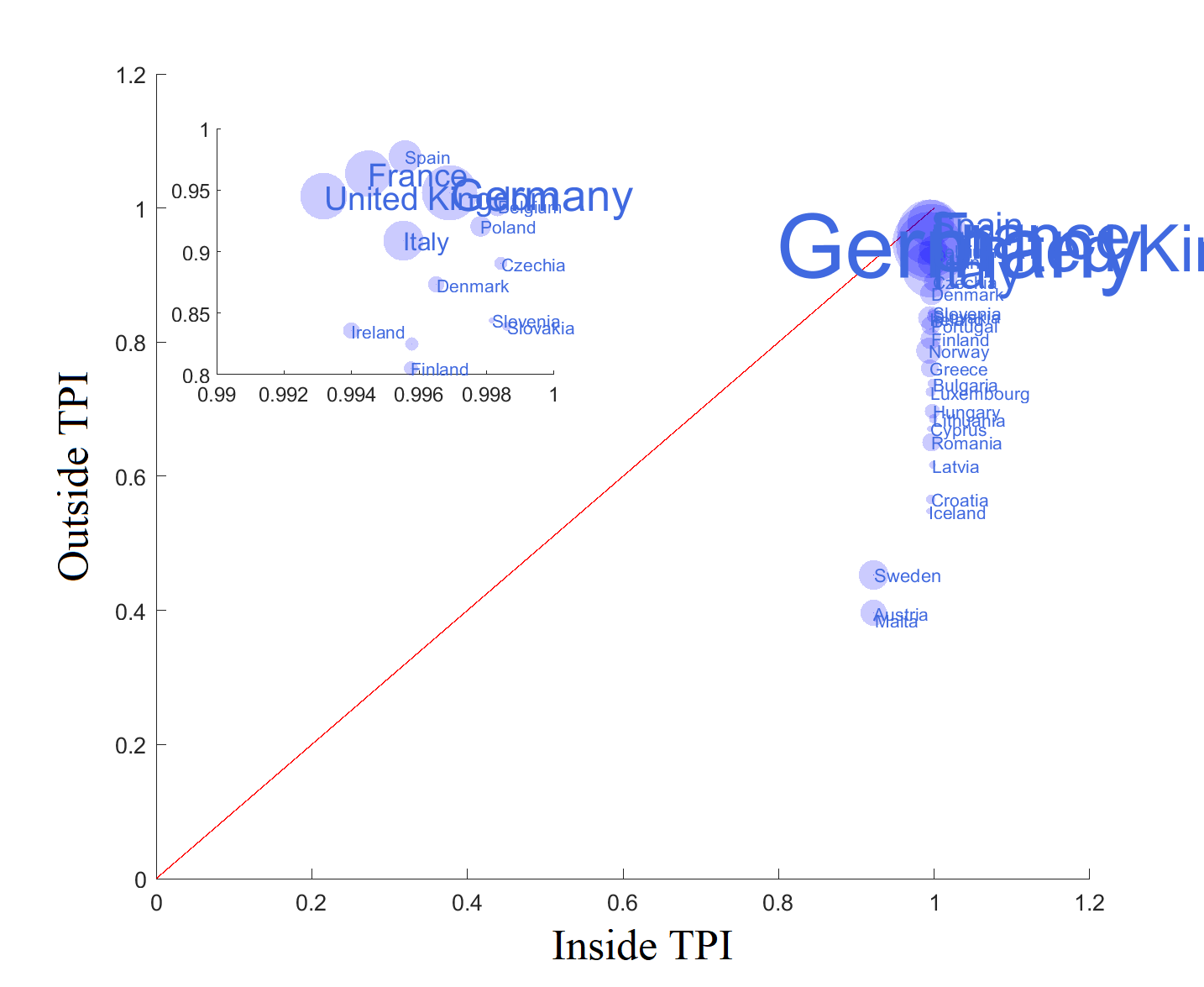}}
    \subfigure[BRI]{\includegraphics[width=.4\textwidth]{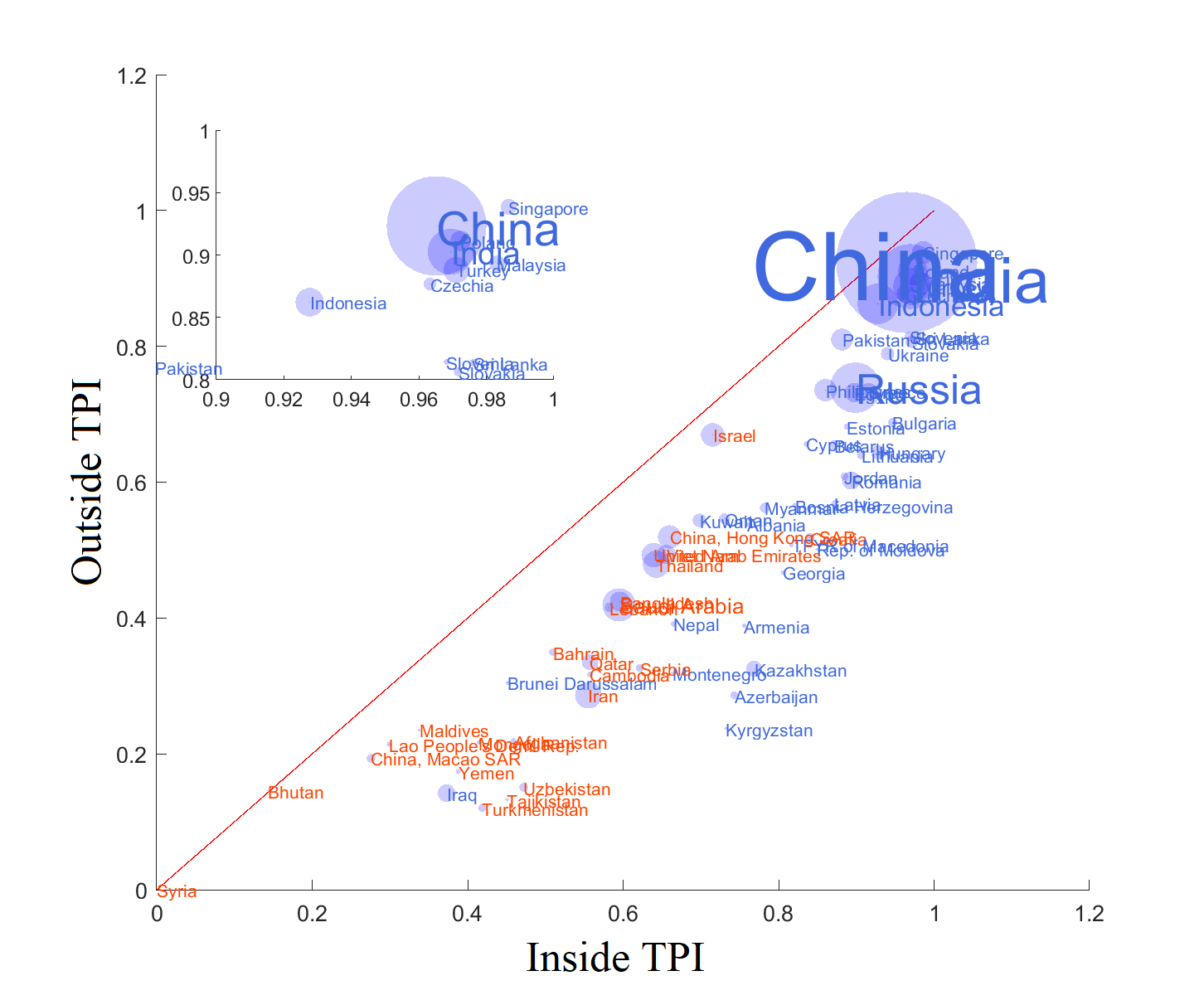}}
    \\
    \subfigure[OAU]{\includegraphics[width=.4\textwidth]{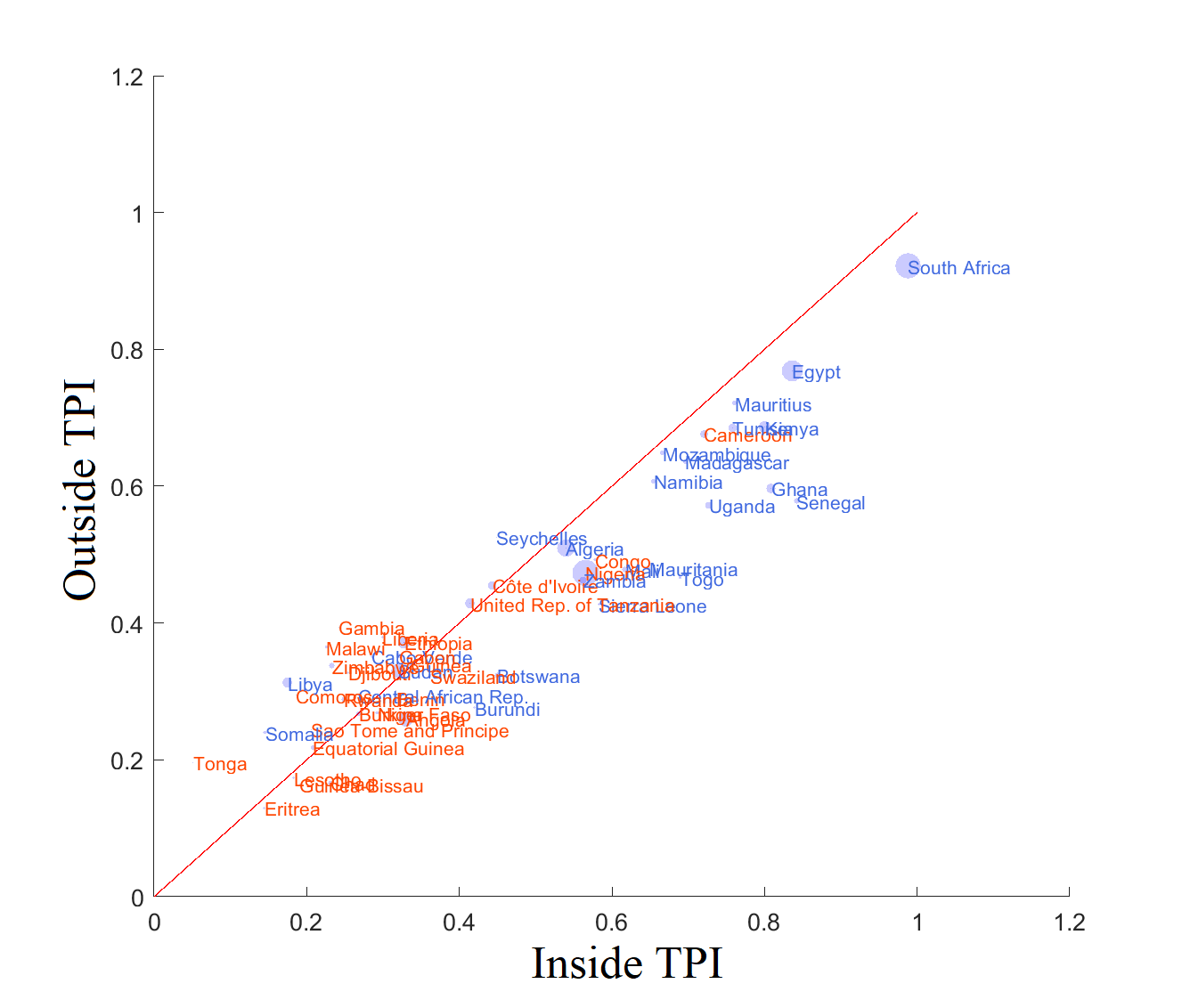}}
    \subfigure[CARIFTA]{\includegraphics[width=.4\textwidth]{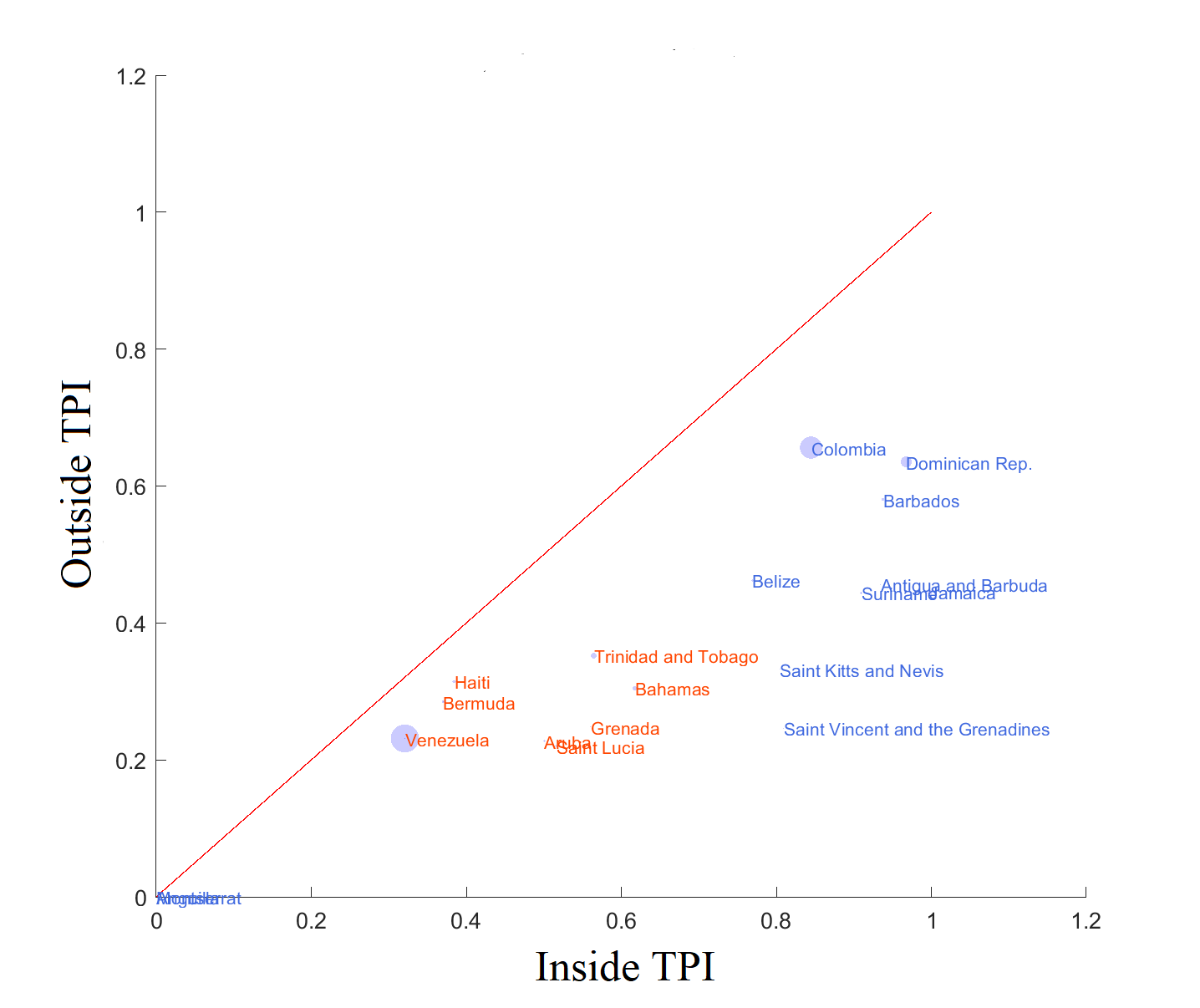}}
    \\
    \subfigure[NAFAT]{\includegraphics[width=.4\textwidth]{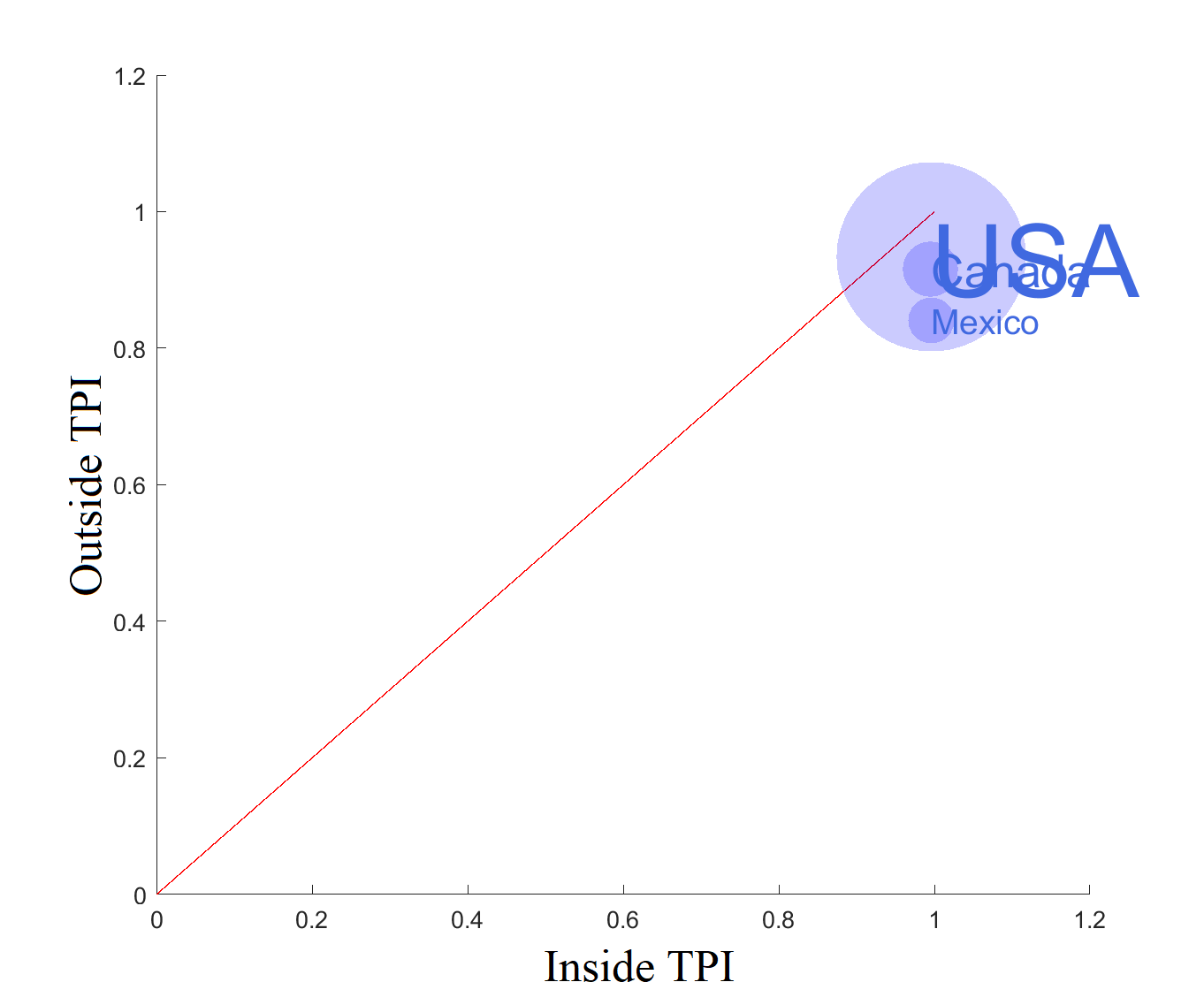}}
    \subfigure[ASEAN]{\includegraphics[width=.4\textwidth]{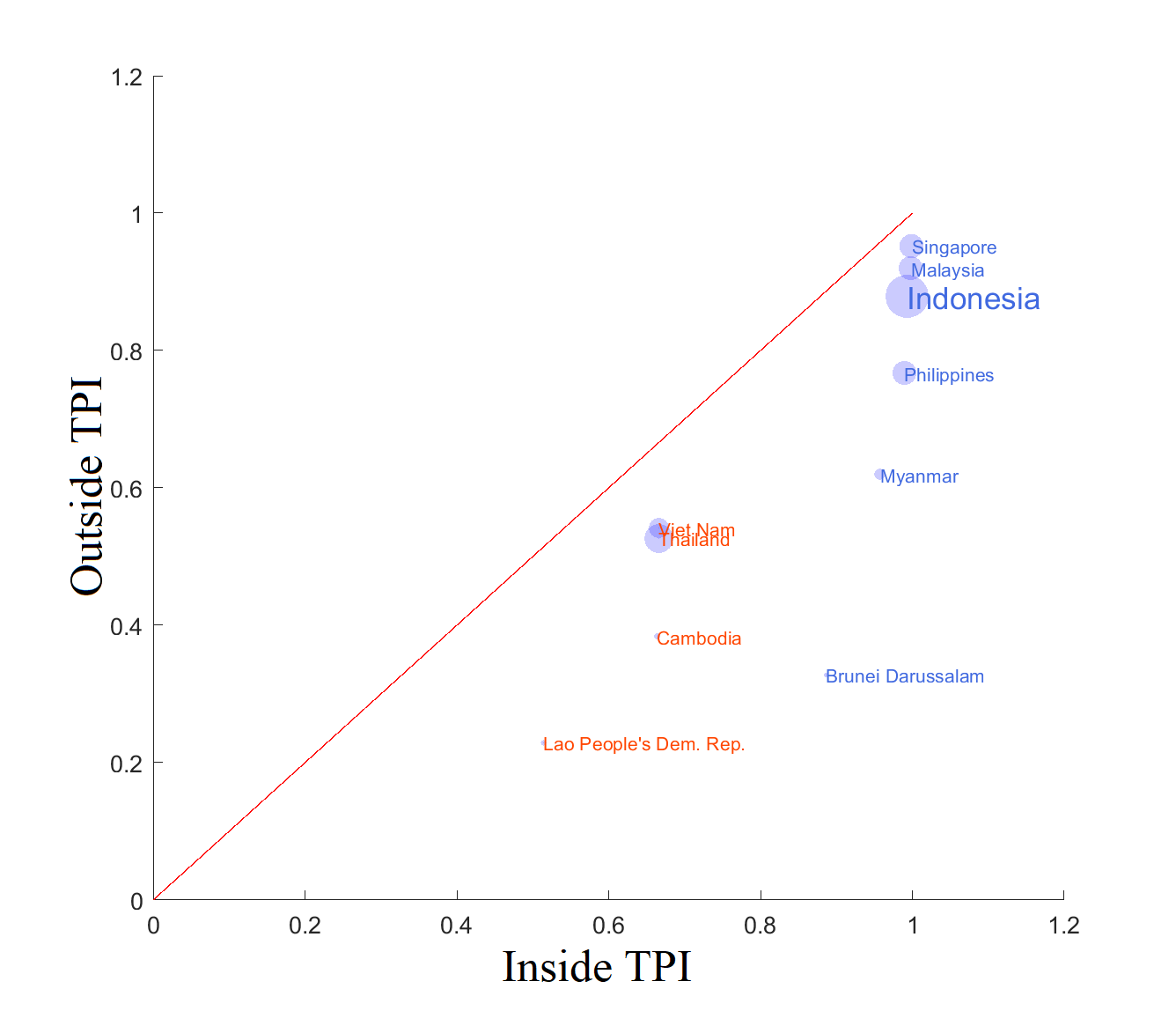}}
   \caption{Trade resistance in six different trade unions for the year 2017. (a) for European Union (EU) countries. (b) for Belt and Road (BRI) countries. (c) for Organization of African Union (OAU) countries. (d) for Caribbean Free Trade Area (CARIFTA) countries. (e) for North American Free Trade Agreement (NAFTA) countries. (f) for Association of Southeast Asian Nations (ASEAN) countries.} 
    \label{fig:TR-6}
\end{figure}

Figure \ref{fig:TR-6} (Appendix) presents some detailed information. The x-coordinate expresses the TPI between a specific country and other countries in the same union, while the y-coordinate expresses the TPI between a specific country and other countries outside the union. The size of dots is proportional to the net trade flow, measured as the absolute value of the difference between exports and imports; a red label means that the country had a trade deficit, while a blue label means a trade surplus flow. Most dots are below the diagonal, which means that the TPIs of most countries inside the union are lower than those outside the union. In addition, countries with surplus trade flow (blue labels) have a higher TPI both inside and outside the union.

\section{Pretreatment of Flow Zero Value}
\label{appendix: zero value}

For the gravity model (equation \ref{grav-r}), $F_{i,j}$ is the trade flow from country $i$ to country $j$; $m_i$ and $m_j$ is the combined size of their economies; $r_{i,j}$ is the trade resistance need to be quantified. It is generally believed that the model cannot describe zero flow because the gravity is universal~\cite{Fatima2019zeroes}, even if the size of two countries is very small and the geographical distance or trade resistance is very large, as long as the volume $m_i \cdot m_j$ is not equal to zero and the resistance $r_{ij}$ are not infinite, the trade flow between them may be very small, but not zero.

\begin{equation}
\label{gravity model}
    \begin{aligned}
    F_{i,j}&\simeq \frac{(m_i\cdot m_j)^\alpha}{r_{i,j}}-\varepsilon_{ij}\\
    &=exp(\alpha \ln(m_i\cdot m_j)-\ln r_{i,j})-\varepsilon_{ij}
    \end{aligned}
\end{equation}

However, the situation of zero-value flow is very common in the empirical data, around 50\% in the global trade network~\cite{helpman2008estimating}, and it creates an additional problem for the log linear form of the gravity equation (including the traditional and structural gravity model in trade studies). In the early studies, some scholars often deal with the zeroes trade observation by truncation method, such as deleting them completely or substitute by small positive constant~\cite{flowerdew1982method,burger2009specification}. It's obviously not rigorous enough~\cite{Fatima2019zeroes}. In reality, the zero-value trade flow is generally considered to be not observable or due to measurement errors from rounding. So stochastic versions of equation are used in empirical studies~\cite{Silva2006The,helpman2008estimating}. Here we can add an error term $\varepsilon_{ij}$, and assume that the error function is positive and obeys lognormal distribution~\cite{Silva2006The}, as $\ln\varepsilon_{ij} \sim N(\mu,\sigma^2)$ in equation \ref{gravity model}.

\begin{equation}\nonumber
\begin{aligned}
E(\varepsilon_{ij})&=e^{\mu^2+\sigma^2/2}\\
Var(\varepsilon_{ij})&=(e^{\sigma^2}-1)e^{2\mu^2+\sigma^2}.
\end{aligned}
\end{equation}

For clarity, we assume $X=\varepsilon_{ij}$, and $Y=X+F_{i,j}$. The probability density function of the random variable $X$ is,

$$ f_X(x)=\left\{
\begin{aligned}
&\frac{1}{\sqrt{2\pi}\sigma x}exp[-\frac{1}{2\sigma^2}(\ln x-\mu)^2]\quad \quad \quad x > 0\\
&\quad \quad \quad \quad \quad 0\quad \quad \quad \quad \quad \quad \quad \quad x \le 0\\
\end{aligned}
\right.
$$

The probability density function of $Y$ is calculated as follows:

\begin{equation}\nonumber
    \begin{aligned}
    &F_Y(y)=P(Y\leq y)=P(F_{i,j}+X\le y)=P(X\le y-F_{i,j})=F_X(y-F_{i,j})\\
    &f_Y(y)=F_Y'(y)=f_X(y-F_{i,j})\times 1
    \end{aligned}
\end{equation}

$$\Rightarrow f_Y(y)=\left\{
\begin{aligned}
&\frac{1}{\sqrt{2\pi}\sigma (y-F_{i,j})}exp[-\frac{1}{2\sigma^2}(\ln(y-F_{i,j})-\mu)^2]\quad \quad \quad y-F_{i,j} > 0\\
&\quad \quad \quad \quad \quad \quad 0 \quad \quad \quad \quad \quad \quad \quad \quad \quad \quad \quad \quad \quad \quad \quad y-F_{i,j} \le 0\\
\end{aligned}
\right.
$$

If we assume that trade resistance is bilateral, then we can simply deduce $r_{i,j}$ for each pair of countries by the least square method with,

\begin{equation}\nonumber
\begin{aligned}
&min(\phi=(F_{i,j}+\varepsilon_{ij}-G\frac{(m_i\cdot m_j)^\alpha}{r_{i,j}})^2+(F_{j,i}+\varepsilon_{ji}-G\frac{(m_i\cdot m_j)^\alpha}{r_{i,j}})^2)\\
 & \frac{\partial \phi}{\partial r_{i,j}}=0 \Rightarrow r_{i,j}^*\simeq \frac{2(m_i\cdot m_j)^\alpha}{F_{i,j}+F_{j,i}+\varepsilon_{ij}+\varepsilon_{ji}}
\end{aligned}
\end{equation}

Different kind of Pseudo Maximum Likelihood (PML) methods are proved to be effective to deal with the zero-valued trade flow and the logarithm transformation~\cite{manning2001estimating,martinez2011log,Silva2006The}. The method in this paper is not exactly the same as the gravity model, and the main different is that we replace the geographical distance with $r_{i,j}$ which needs to be quantified. So we use the idea of PML, but improve the likelihood function here. Then, we maximize the probability $E(Y)=E(X)+F_{i,j}$, with the defined likelihood function,

\begin{equation}\nonumber
\begin{aligned}
L&=\prod_{i,j}p(E(X)+F_{i,j}|\mu,\sigma)=\prod_{i,j}p(E(G\frac{(m_i\cdot m_j)^\alpha}{r_{i,j}})|\mu,\sigma)\\
&\simeq\prod_{i,j}p(\frac{2(m_i\cdot m_j)^\alpha}{F_{i,j}+F_{j,i}+2E(\varepsilon_{ij})}|\mu,\sigma)
\end{aligned}
\end{equation}

With the method of maximum likelihood estimation, we can optimize the parameters $\mu$ and $\sigma$ to get the $max_{\mu,\sigma} (L)$, which make $E(Y)=E(X)+F_{i,j}$ the most likely to occur in reality.

\begin{figure}
    \centering
    \includegraphics[width=.7\textwidth]{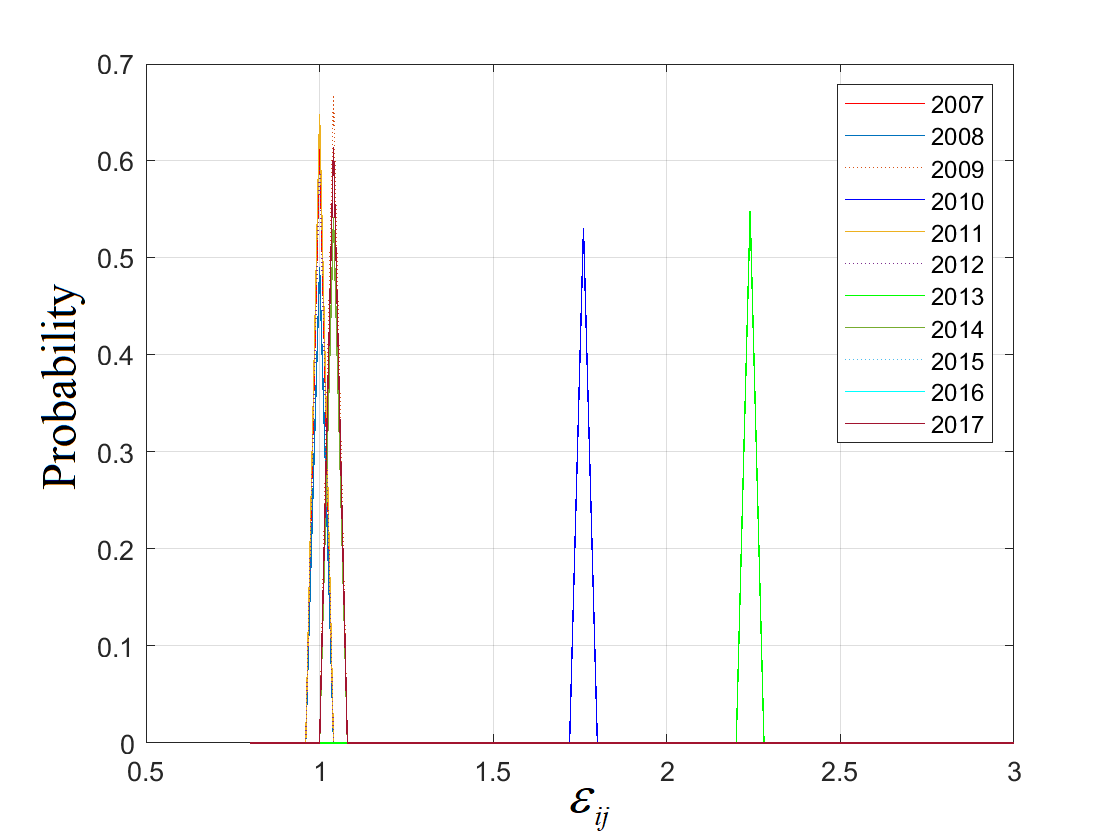}
    \caption{The random error obeys lognormal distribution during 2007-2017.}
    \label{fig:distribution of error}
\end{figure}

The optimized parameters are listed in Table \ref{delta}, and Figure \ref{fig:distribution of error} shows the distribution of random error $\varepsilon_{ij}$ during 2007-2017. It can be seen that the mean value of random variables is basically around 1-2, and the variance is relatively small, which conforms to the basic assumption of statistical error in trade flows.

\begin{table}
\centering 
\begin{tabular}{c|c|c|c}
 \hline
  \hline
 & $\mu$ & $\sigma$ & $E(\varepsilon_{ij})$\\
 \hline
2007 & 0.00694  & 0.00050  & 1.00697 \\
2008 & 0.00228  & 0.00020  & 1.00229 \\
2009 & 0.02364  & 0.00047  & 1.02392 \\
2010 & 0.56409  & 0.00027  & 1.75785 \\
2011 & 0.00339  & 0.00024  & 1.00340 \\
2012 & 0.01529  & 0.00072  & 1.01540 \\
2013 & 0.81314  & 0.00018  & 2.25498 \\
2014 & 0.05607  & 0.00061  & 1.05767 \\
2015 & 0.40263  & 0.00017  & 1.49575 \\
2016 & 0.31945  & 0.00028  & 1.37637 \\
2017 & 0.02362  & 0.00047  & 1.02390 \\
\hline
\end{tabular}
	\caption{Optimized Parameters}
		\label{delta} 
\end{table}

\end{document}